\documentclass[fleqn,usenatbib]{rasti}
\usepackage{newtxtext,newtxmath}

\usepackage[T1]{fontenc}

\DeclareRobustCommand{\VAN}[3]{#2}
\let\VANthebibliography\thebibliography
\def\thebibliography{\DeclareRobustCommand{\VAN}[3]{##3}\VANthebibliography}

\usepackage{graphicx}	
\usepackage{float}
\graphicspath{Figures/}
\usepackage[color=blue!20!white, obeyFinal]{todonotes}

\title[Learned Interferometric Imaging for SPIDER]{Learned Interferometric Imaging for the SPIDER Instrument}

\author[M. Mars et al.]{
Matthijs Mars,$^{1}$\thanks{E-mail: matthijs.mars.20@ucl.ac.uk}
Marta M.~Betcke,$^{2}$
Jason D.~McEwen$^{1, 3}$
\\
$^{1}$Mullard Space Science Laboratory (MSSL), University College London (UCL), Dorking RH5 6NT, UK\\
$^{2}$Department of Computer Science, University College London (UCL), London WC1E 6BT, UK \\
$^{3}$Alan Turing Institute, London NW1 2DB, UK
}

\date{Accepted 2023 November 1. Received 2023 September 29; in original form 2023 February 20}

\pubyear{2023}

\usepackage{xcolor}

\hypersetup{
  colorlinks=true,
}

\begin{document}
\label{firstpage}
\pagerange{\pageref{firstpage}--\pageref{lastpage}}
\maketitle

\begin{abstract}
  The Segmented Planar Imaging Detector for Electro-Optical Reconnaissance (SPIDER) is an optical interferometric imaging device that aims to offer an alternative to the large space telescope designs of today with reduced size, weight and power consumption. This is achieved through interferometric imaging. State-of-the-art methods for reconstructing images from interferometric measurements adopt proximal optimization techniques, which are computationally expensive and require handcrafted priors. In this work we present two data-driven approaches for reconstructing images from measurements made by the SPIDER instrument. These approaches use deep learning to learn prior information from training data, increasing the reconstruction quality, and significantly reducing the computation time required to recover images by orders of magnitude.  Reconstruction time is reduced to ${\sim} 10$ milliseconds, opening up the possibility of real-time imaging with SPIDER for the first time. Furthermore, we show that these methods can also be applied in domains where training data is scarce, such as astronomical imaging, by leveraging transfer learning from domains where plenty of training data are available.
\end{abstract}



\section{Introduction} \label{sec:introduction}
The Segmented Planar Imaging Detector for Electro-Optical Reconnaissance \citep[SPIDER;][]{kendrickFlatPanelSpaceBasedSpace2013,duncanSPIDERNextGeneration2015} instrument is an alternative electro-optical imaging device to current space telescopes like the Hubble Space Telescope or the James Webb Space Telescope.
While traditional electro-optical telescopes require large optics, housings, and thermal controls for the optics to attain precise measurements, SPIDER aims to decrease the volume, mass, and cost of electro-optical imagers by replacing the traditional mirrors with arrays of lenslets to gather interferometric measurements.
The light captured by these lenslets is processed using photonic integrated circuit (PIC) chips into interferometric measurements. Typical interferometers still require large components to make measurements, yet by processing the light from the lenslets on an integrated chip the size of the instrument can be reduced significantly.
The SPIDER instrument is designed to be a cheaper and lighter alternative for instruments that can be used for both Earth observation or astronomical research.

\begin{figure}
  \centering
  \includegraphics[width=\columnwidth]{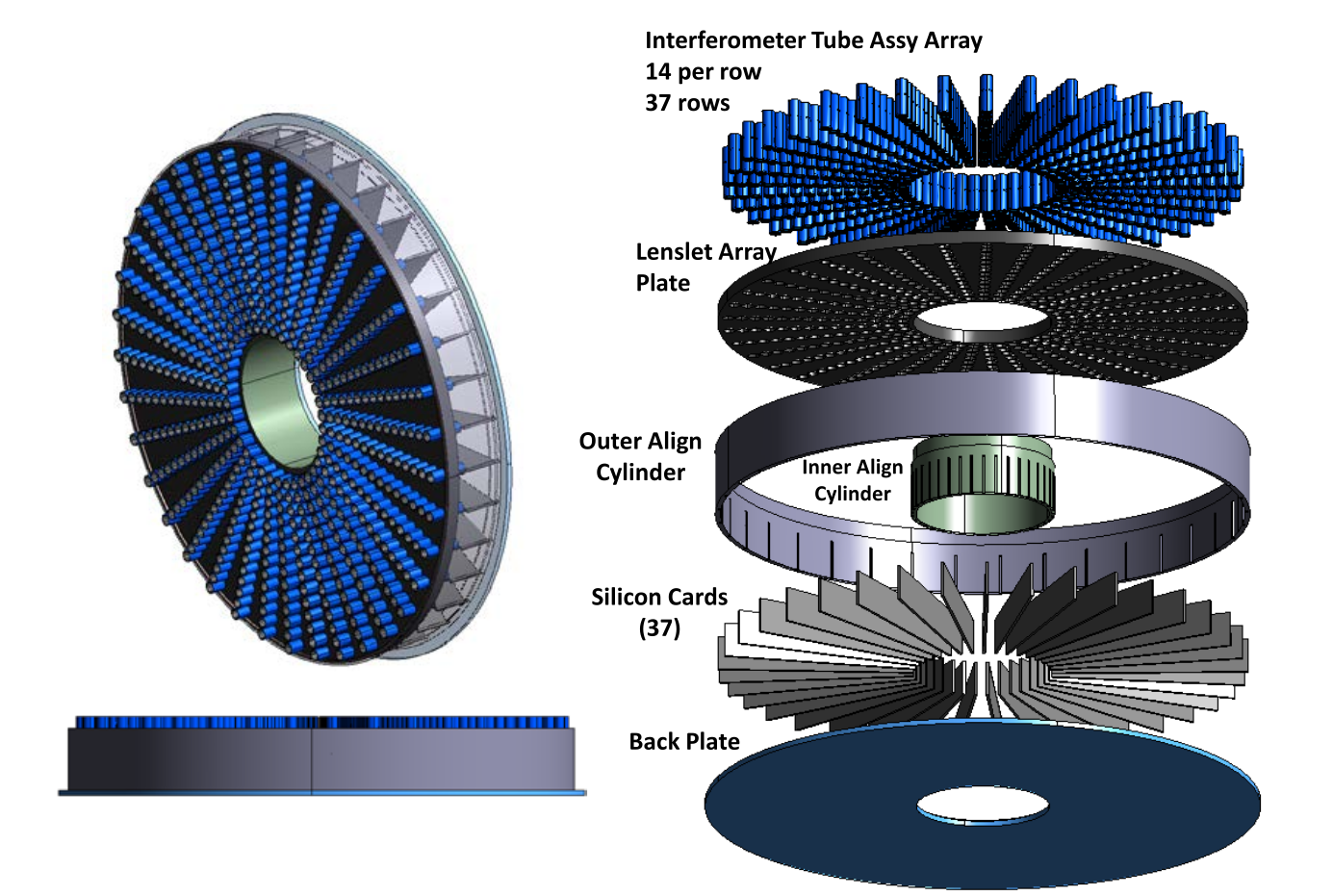}
  \caption{A diagram of a SPIDER instrument design proposed with 37 PICs with lenslets attached to them. Image credit: \citet{kendrickFlatPanelSpaceBasedSpace2013}. }
  \label{fig:spider-instrument}
\end{figure}

The concept design for the SPIDER instrument proposed by \citet{kendrickFlatPanelSpaceBasedSpace2013} and \citet{duncanSPIDERNextGeneration2015} uses 37 PICs mounted at different angles to form a planar disk. Interferometric measurements are acquired on each of the PICs, and by combining the measurements taken at each of the different angles, the telescope creates a large synthetic aperture with the diameter of the aperture equal to the largest spacing of lenslets on the PIC. A diagram of their design can be found in Figure~\ref{fig:spider-instrument}.

Beyond the initial concept, more efficient designs that make better use of the baselines \citep{liuSystemDesignOptical2018, liuImageReconstructionEmerging2019} and improved lenslet arrangements to increase reconstructions quality \citep{lvSystemDesignImproved2020,huOptimalDesignSegmented2021} have been investigated. Furthermore, the PIC reconstruction capabilities were found to match the theoretical predictions of the concept design well \citep{chuNumericalSimulationOptimal2017}. Tests of the PICs show that it is possible to measure both the amplitude and phase of the incoming light as predicted \citep[][]{suExperimentalDemonstrationInterferometric2017} and that these measurements can be used for image reconstruction \citep[][]{badhamPhotonicIntegratedCircuitbased2017,suInterferometricImagingUsing2018}.

Since SPIDER measures both the phase and the amplitude of the interferometric signals, the measurement process is analogous to that of radio interferometers. Interferometric imaging techniques developed for radio interferometry can thus be adapted and repurposed to recover images from the raw data acquired by the SPIDER instrument \citep{pratleySparseImageReconstruction2021}.  Radio interferometry and aperture synthesis have played an important role in pushing the boundaries of astronomical research in the radio frequency regime, where acquiring high resolution images is otherwise difficult because of the relatively large wavelengths considered. By combining measurements from pairs of radio telescopes, spatial frequency information can be retrieved from the observed sky. Each pair of telescopes forms a baseline and measures a so-called visibility corresponding to a Fourier coefficient of the sky brightness distribution. By sampling the Fourier plane with these measurements, an observation with a uniform aperture is approximated using aperture synthesis.  Since the telescope acquires only a finite number of measurements and since the Fourier plane is not sampled uniformly, the problem is ill-posed and an accurate image cannot be reconstructed simply by inverting the Fourier transform. Instead, when inverting the measuring process directly as described a so-called dirty image is obtained: a reconstruction of the sky brightness convolved with the point spread function (PSF) of the telescope configuration. In order to recover an accurate image of the sky brightness, techniques to regularized inverse problems are typically considered.

One method for solving this inverse problem is variational regularization. Variational regularization techniques combine a data fidelity component relative to a forward model with a regularizer in order to constrain and stabilize the problem. Regularizers for these algorithms encode prior information about the images. An example of modern model-based iterative methods applied to radio interferometry can be found in \citet{pratleyRobustSparseImage2018}, which adopt convex optimization techniques to solve the variational regularization optimization problem.  An application of these techniques to SPIDER imaging can be found in \citet{pratleySparseImageReconstruction2021}. While these techniques can be distributed to reduce the reconstruction time \citep[][]{pratleyDistributedParallelSparse2019}, they are computationally expensive as they evaluate the measurement operator that models the instrument at each iteration of the optimization process. Besides the computational cost of these algorithms, they also adopt handcrafted priors (e.g.\ $\ell_1$ sparsity in a wavelet representation) that, while general, are not tailored to the images of interest.

An alternative to model-based algorithms, such as variational regularization, are data-driven algorithms that learn the prior information from training data. Since the prior is implicitly specified by the training data, data-driven methods are typically able to achieve higher reconstruction quality than by using handcrafted priors, provided the training distribution matches the target distribution well.  Learned methods that are completely independent of the measurement operator and attempt to learn a direct mapping from measurements to target image, however, are generally not effective.  Thus, learned methods typically combine some model-based information, like the measurement operator, with prior information learned from training data.
Learned data-driven approaches broadly fall into three categories, differentiated by the degree to which model-based information such as the measurement operator is encoded or leveraged.
\textit{Learned regularization methods} \citep[e.g.][]{lunzAdversarialRegularizersInverse2018,liNETTSolvingInverse2020} are iterative in nature and make full use of the measurement operator in each iteration; consequently, they achieve excellent reconstruction quality by exploiting full knowledge of the measurement operator, along with learned prior information, but are generally highly computationally demanding.
\textit{Learned sequential methods} \citep[e.g.][]{jinDeepConvolutionalNeural2017}, on the other hand, simply pre- and/or post-process data with learned models in observation and/or image space; consequently, while they nevertheless provide good quality reconstructions they are limited by the fact that they only evaluate the measurement operator very few times (sometimes just once). This however does make these methods substantially more efficient computationally.
\textit{Learned iterative methods} \citep[e.g.][]{adlerSolvingIllposedInverse2017}, also called unrolled methods, provide a balance by designing a learned model that attempts to unroll a small number of iterations of iterative approaches, encoding the measurement operator into the model; consequently, they typically achieve superior reconstruction quality to learned sequential methods exploiting greater knowledge of the measurement operator and are more computationally efficient than learned regularization methods.

Many such methods were originally proposed in the medical imaging context (CT, MRI, PAT) and are based on deep learning approaches \citep[e.g.][]{adlerSolvingIllposedInverse2017,adlerLearnedPrimaldualReconstruction2018,arridgeSolvingInverseProblems2019}.
%
In the field of radio astronomy learned image reconstruction techniques were first considered by McEwen \& Allam Jr \citep[]{allamjrRadioInterferometricImage2016}, where super resolution convolutional neural networks \citep[SRCNN;][]{dongImageSuperResolutionUsing2016} were applied to post-process dirty radio interferometric images. These networks were considered for cases where the exact telescope PSF is known, as well as for the PSF-unaware case, highlighting the potential of learning a generalized network that works with unseen telescope configurations. Although, performance was relatively poor for this rudimentary approach.  More recent applications of learned post-processing methods in radio astronomy have been considered using learned denoisers \citep{terrisDeepPostProcessingSparse2019}, convolutional autoencoders \citep[][]{ghellerConvolutionalDeepDenoising2021}, and super resolution networks \citep[][]{connorDeepRadiointerferometricImaging2022}.  The use of plug-and-play \citep[PnP;][]{venkatakrishnanPlugandPlayPriorsModel2013} denoisers within iterative reconstruction methods has also been considered for radio interferometric imaging \citep{terrisImageReconstructionAlgorithms2022}, although such approaches still require many evaluations of the measurement operator during imaging.

While we draw inspiration from these prior works on learned imaging for interferometry, in this paper we develop new learned imaging techniques, specifically targeting the SPIDER instrument.  Our primary goal is reducing computational cost, while of course still ensuring high quality reconstructions.  Reducing the computational cost of recovering images from raw SPIDER measurements would open up the possibility of real-time imaging with SPIDER, which would afford numerous new applications.  Consequently, we develop a learned sequential method that requires only a single evaluation of the measurement operator.  While reconstruction quality for this method is similar to current state-of-the-art variational regularization techniques, computational time is orders of magnitude faster.  We also develop a learned iterative method, trading off a small increase in computational time compared to the learned sequential method, but that nevertheless remains orders of magnitude faster than traditional approaches, while achieving a further improvement in reconstruction quality.


The remainder of the paper is structured as follows. Section~\ref{sec:imaging} introduces the measurement process of interferometric imagers and the particular imaging configuration of the SPIDER instrument, as well as discussing the inverse imaging problem. Variational and learned approaches for the inverse imaging problems are reviewed in Section~\ref{sec:inverse-problems}. In Section~\ref{sec:modelling-spider} we present two approaches to model the SPIDER instrument: alongside the standard non-uniform Fourier transform approach considered previously \citep{pratleySparseImageReconstruction2021}, we also introduce a new modelling approach for SPIDER based on the Radon transform. In Section~\ref{sec:solving-approaches} we propose our learned methods for reconstruction from interferometric measurements. Section~\ref{sec:results} evaluates the performance of the reconstruction methods when applied to natural images, their robustness to additional noise, and generalization potential of these methods to smaller datasets of galaxy or satellite images.   Finally, concluding remarks are made in Section~\ref{sec:conclusion}.

\section{SPIDER Instrument} \label{sec:imaging}
The SPIDER instrument measures incoming light through pairs of separated lenses and combines it in waveguides on a PIC chip to form interferometric baselines. All the operations to make the interferometric measurements are performed on the chip, resulting in a small form-factor for the system. The concept design introduced in \citet{kendrickFlatPanelSpaceBasedSpace2013} uses a linear array of lenslets attached to one PIC chip to measure several baselines using a 1D interferometer. Several of these so-called spokes are then mounted radially resulting in a two-dimensional sampling pattern. While in typical interferometers the information of all receivers can be combined into interferometric measurements, resulting in a total of $N(N-1)/2$ measurements for $N$ lenslets, in the SPIDER design lenslets can only be combined within one PIC chip, resulting $N/2$ baselines per PIC module. To increase the number of baselines gathered from one PIC card, different wavelengths of light are measured. Because the spatial frequency measured depends on the separation of the lenslets and the wavelength of the light, the number of baselines is multiplied by the number of wavelength bins observed. The technique of incorporating spectral information from measurements at different wavelengths in a single reconstruction is called multi-frequency synthesis and is common in radio interferometry \citep{saultMultiFrequencySynthesis1999}.

\begin{figure}
    \centering
    \includegraphics[width=\columnwidth]{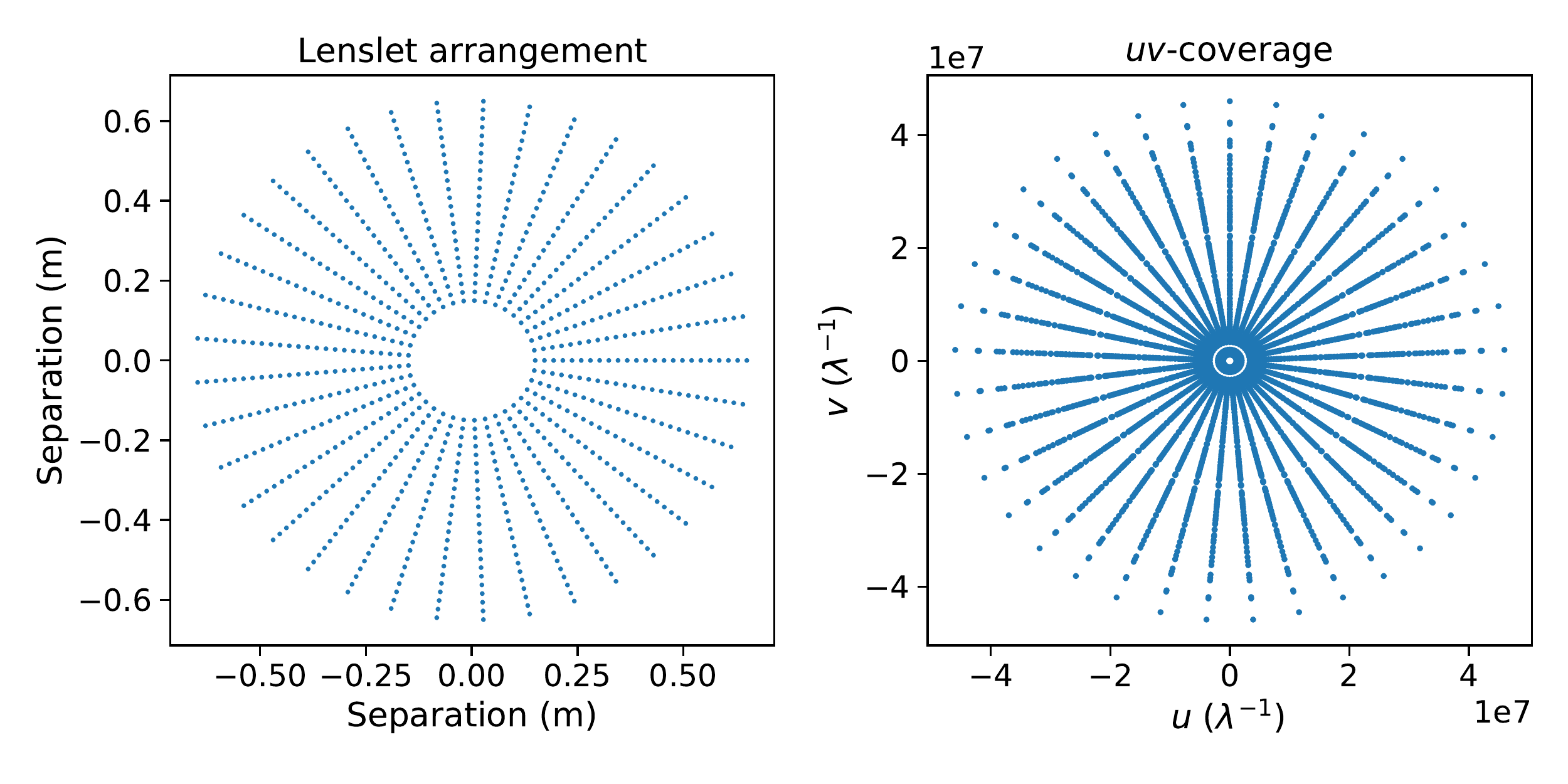}
    \caption{\emph{(Left)} The physical locations of the lenslets of the 2D interferometer as detailed in the design proposed in \citet{kendrickFlatPanelSpaceBasedSpace2013}. The lenslets on each of the radial spokes are mounted to their respective PIC. \emph{(Right)} The measured baselines of the interferometric measurements between pairs of lenslets on the PICs of the SPIDER instrument. The amount of baselines is increased by measuring at different spectral frequencies. Note that the measurements all lie in the same direction as the directions of the spokes, since measurements are only made using pairs of lenslets on 1 PIC.  }
    \label{fig:spider-config}
\end{figure}

By mounting multiple PIC chips as radial spokes on a disc, a 2D interferometer is created. In doing so a radial sampling profile is created for the $uv$-plane. In the concept of \citet{kendrickFlatPanelSpaceBasedSpace2013}, 37 spokes of PICs with 24 lenslets each are used. The measured light frequency spectrum is from $500\text{nm}$ to $900\text{nm}$ with 10 spectral bins. This results in $120$ baselines per spoke and a total of $4440$ measured Fourier components. The parameters for the configuration used by \citet{kendrickFlatPanelSpaceBasedSpace2013} can be found in Table~\ref{tab:spider-config} and the lenslet layout and the $uv$-sampling are shown in Figure~\ref{fig:spider-config}.

\begin{table}
    \centering
    \caption{The parameters of the SPIDER concept design proposed in \citet{kendrickFlatPanelSpaceBasedSpace2013}.}
    \label{tab:spider-config}
    \begin{tabular}{l|c}\hline
        Parameter                       & Value      \\ \hline
        Spectral range                  & 500-900 nm \\
        Lenslet Diameter                & 8.75 mm    \\
        Longest Baseline                & 0.5 m      \\
        Number of Lenslets per PIC card & 24         \\
        Number of PIC cards             & 37         \\
        Number of Spectral Bins         & 10         \\ \hline
    \end{tabular}
\end{table}

The interferometric measurements at the baseline (spatial) frequency $\boldsymbol{\xi} = (u,v)$ represent samples of the 2D Fourier transform of the image $f(\boldsymbol{\chi})$, with spatial coordinate $\boldsymbol{\chi}$, as given by the van Cittert-Zernike theorem \citep{zernikeConceptDegreeCoherence1938}:
\begin{equation}\label{eq:visibility}
    \hat{f}(\boldsymbol{\xi}) = \int_{-\infty}^{\infty} \int_{-\infty}^{\infty} f(\boldsymbol{\chi}) \ \mathrm{e}^{-\mathrm{i\,} 2 \pi  \boldsymbol{\chi} \cdot \boldsymbol{\xi} } \mathrm{d}\boldsymbol{\chi},
\end{equation}
which is a continuous unitary Fourier transform of the signal. The baselines are determined by the spatial distance between the lenslets on the PIC with respect to the observed wavelength of the light.
To recover an image from the Fourier measurements one needs the inverse of Equation~\ref{eq:visibility},
\begin{equation}\label{eq:FT-inverse}
    f(\boldsymbol{\chi}) = \int_{-\infty}^{\infty} \int_{-\infty}^{\infty} \hat{f}(\boldsymbol{\xi}) \  \mathrm{e}^{\mathrm{i\,} 2 \pi  \boldsymbol{\chi} \cdot \boldsymbol{\xi} } \mathrm{d}\boldsymbol{\xi},
\end{equation}
which is only possible in theory when one has complete knowledge of the continuous Fourier representation $\hat{f}(\boldsymbol{\xi})$.  In practice this is not the case.

The approaches for modelling the SPIDER instrument introduced in Section~\ref{sec:modelling-spider} require an (upsampled) discrete Fourier transform (DFT).  The unitary DFT for an $N_1 \times N_2$ image is an operator mapping from $\mathbb{R}^{N_1\times N_2} \rightarrow \mathbb{C}^{M_1 \times M_2}$, we henceforth use the same $\hat f$ notation for the continuous and discrete Fourier transform as the meaning will be clear from the context. The unitary DFT is defined by
\begin{equation}\label{eq:DFT-forward}
    \hat{f}(\boldsymbol{\xi}_{kl}) = \frac{1}{\sqrt{N_1 N_2}} \sum_{{i}=1}^{N_1} \sum_{{j} =1}^{N_2} f(\boldsymbol{\chi}_{{ij}}) \  \mathrm{e}^{-\mathrm{i\,} \boldsymbol{\chi}_{{ij}} \cdot \boldsymbol{\xi}_{kl}},
\end{equation}
and the unitary inverse DFT by
\begin{equation}\label{eq:DFT-inverse}
    f(\boldsymbol{\chi}_{ij}) =  \frac{1}{\sqrt{M_1 M_2}}\sum_{k=1}^{M_1} \sum_{l=1}^{M_2} \hat{f}(\boldsymbol{\xi}_{kl}) \ \mathrm{e}^{\mathrm{i\,} \boldsymbol{\chi}_{ij} \cdot \boldsymbol{\xi}_{kl}}.
\end{equation}

The interferometric measurement process (Equation~\ref{eq:visibility}) can be written in the compact, discretized form
\begin{equation}
    \label{eq:inverse-problem}
    \boldsymbol{y} = \boldsymbol{\Phi}\boldsymbol{x} + \boldsymbol{n},
\end{equation}
where the linear measurement operator $\boldsymbol{\Phi}: X \rightarrow Y$, maps the unknown image $\boldsymbol{x} \in X \subset \mathbb{R}^N$ to the set of noisy measurements $\boldsymbol{y} \in Y \subset \mathbb{C}^M$ (here Fourier components of the weighted sky-brightness), with $\boldsymbol{n} \in Y \subset \mathbb{C}^M$ some type of measurement noise. The interferometric measurement operator corresponds to a non-uniformly sampled Fourier transform.

The limited number of lenslets of the telescope results in a limited sampling of the $uv$-plane. Since the Fourier domain is sampled incompletely, the measurement operator is ill-posed, and it cannot simply be inverted to find a solution to the inverse problem. Because of the ill-posedness of the operator, prior information on the solution is needed to regularize the inversion. Therefore, approaches such as sparse regularization are needed to stably recover a solution for the reconstruction. Underlying the sparse regularization is the idea that natural signals (e.g. astronomical images) are sparse or compressible in a suitable basis or frame (e.g. wavelet bases).
\section{Inverse Imaging Approaches}\label{sec:inverse-problems}

We briefly recall the state-of-the-art variational techniques that are applied to solve the interferometric imaging problem and how these methods can be enhanced or replaced by using learned approaches that make use of deep learning.

\subsection{Variational regularization}

To recover a solution to the inverse problem stated in Equation~\ref{eq:inverse-problem}, we find a solution for $\boldsymbol{x}$ that trades off matching the data and the prior information encoded in the regularizer.  Variational regularization approaches do this by posing an appropriate optimization problem, which is then often solved by proximal optimization algorithms.

\subsubsection{Optimization problems}

Trading off data fidelity and prior information can be achieved by obtaining a solution to the following minimization problem:
\begin{equation}
	\label{eq:variational-regularisation}
	\boldsymbol{x}^\star = \underset{\boldsymbol{x} \in X}{\text{arg min}} \quad \mathcal{L}( \boldsymbol{\Phi} \boldsymbol{x}, \boldsymbol{y}) + \lambda \mathcal{S}(\boldsymbol{x}),
\end{equation}
where $\lambda$ is the regularization parameter, and $\mathcal{L}( \boldsymbol{\Phi} \boldsymbol{x}, \boldsymbol{y})$ and $\mathcal{S}(\boldsymbol{x})$ are a data fidelity and regularization functional respectively.

A common optimization strategy is to use a regularizer that promotes sparsity in a particular frame. The signal $\boldsymbol{x} \in \mathbb{R}^N$ can be represented in a dictionary $\boldsymbol{\Psi} \in \mathbb{R}^{N \times D}$ by $\boldsymbol{x}  = \boldsymbol{\Psi} \boldsymbol{\alpha}$, where $\boldsymbol{\alpha} \in \mathbb{R}^D$. Natural signals typically exhibit stronger sparsity if the dictionary is redundant, i.e.  $D > N$ \citep{gribonvalSparseRepresentationsUnions2003,bobinMorphologicalComponentAnalysis2007,starckSparseImageSignal2010}. It is therefore often beneficial to use dictionaries which are a concatenation of different basis of frames, e.g.\ Dirac and Daubechies wavelets \citep[][]{carrilloSparsityAveragingReweighted2012,carrilloSparsityAveragingCompressive2013}.

The minimization problem can be expressed in the synthesis formulation, where the image $\boldsymbol{x}^\star$ is synthesized from dictionary elements by $\boldsymbol{x}^\star = \boldsymbol{\Psi} \boldsymbol{\alpha}^\star$. Alternatively, the problem is expressed in the so-called analysis setting where the image $\boldsymbol{x}^\star$ is recovered directly while promoting the sparsity of $\boldsymbol{\Psi}^* \boldsymbol{x}^\star$. If $\boldsymbol{\Psi}$ is a tight frame 
the adjoint $\boldsymbol{\Psi}^*$ is equal to the self-inverse $\boldsymbol{\Psi}^\dagger$ and the two formulations are equivalent. However, in a general overcomplete dictionary (such as a concatenation of two bases) the analysis formulation is often found to yield superior reconstruction quality  \citep[e.g.][]{carrilloSparsityAveragingReweighted2012,carrilloSparsityAveragingCompressive2013}.

The data fidelity term is typically the $\ell_2\text{-norm}$ of the residuals (which can also be interpreted as the log-likelihood for the case of Gaussian noise). The optimization functional in the unconstrained, analysis setting then reads
\begin{equation}
	\label{eq:unconstrained}
	\boldsymbol{x}^\star = \underset{\boldsymbol{x} \in X}{\text{arg min}} \quad  \|\boldsymbol{\Phi} \boldsymbol{x} - \boldsymbol{y} \|_{\ell_2}^2 + \lambda \| \boldsymbol{\Psi}^\dagger \boldsymbol{x} \|_{\ell_1},
\end{equation}
assuming identity covariance (typically measurements are weighted to have unit variance; as discussed in Section~\ref{sec:nufft}).

In the unconstrained setting the optimization depends on a good choice of the hyperparameter $\lambda$ to find a good balance between the data fidelity and the sparsity of the signal. The problem can also be formulated as a constrained optimization problem
\begin{equation}
	\label{eq:constrained-analysis}
	\boldsymbol{x}^\star = \underset{\boldsymbol{x} \in X}{\text{arg min}}  \quad \| \boldsymbol{\Psi}^\dagger \boldsymbol{x} \|_{\ell_1}, \quad \text{s.t.} \quad  \| \Phi \boldsymbol{x} - \boldsymbol{y} \|_{\ell_2} < \epsilon ,
\end{equation}
where we seek the most sparse $\boldsymbol{\Psi}^\dagger \boldsymbol{x}$ that satisfies the constraint with some $\epsilon > 0$ on the data misfit. The hyperparameter $\epsilon$ can be estimated from an estate of the noise level \citep{pratleyRobustSparseImage2018}.

\subsubsection{Proximal optimization algorithms}\label{sec:prox-algorithms}

The optimization problems described in the previous subsection can often be solved by proximal optimization algorithms that leverage a proximity (or proximal) operator. The proximity operator of a (proper, semi-continuous) convex function $\lambda h$ (with $\lambda > 0$) maps $\boldsymbol{v} \in \mathbb{R}^N$ to a unique solution to the (strongly convex) minimization problem
\begin{equation}\label{eq:proximal-operator}
	\text{prox}_{\lambda h}(\boldsymbol{v}) = \underset{\boldsymbol{x}\in X}{\text{argmin}} \  \lambda h (\boldsymbol{x}) + \frac{1}{2} \|\boldsymbol{x} - \boldsymbol{v}\|^2_{\ell_2} .
\end{equation}
The parameter $\lambda$ sets  the balance between the squared $\ell_2$-distance to $\boldsymbol{v}$ and the value of $h$. Many common proximal operators admit an analytical solution or at least a linear time iterative solution. The fixed point of a proximal operator is the global minimum of $h$ \citep[][]{boydConvexOptimization2004,combettesProximalSplittingMethods2011}.

Proximal splitting methods use proximal operators to estimate the solution to the inverse problem by splitting the objective function in separate steps for the different optimization functionals. A review of different proximal splitting methods can be found in \citet{combettesProximalSplittingMethods2011}. The simplest proximal splitting algorithm is the proximal gradient method, which consists of a gradient update step, followed a proximal update step:
\begin{equation}\label{eq:forward-backward}
	\boldsymbol{x}_{i+1} = \text{prox}_{\lambda \mathcal{S}} \left(\boldsymbol{x}_{i} - \nabla \mathcal{L} (\Phi \boldsymbol{x}_{i} ) \right).
\end{equation}
In this manner sparsity-promoting priors $\mathcal{S}(\boldsymbol{x})$ that are not differentiable can be supported (e.g.\ $\ell_1$ sparsity in a wavelet basis).

Proximal splitting methods have found numerous applications to radio interferometric imaging problems in astronomy: Douglas-Rachford splitting \citep[e.g.][]{carrilloSparsityAveragingReweighted2012, carrilloSparsityAveragingCompressive2013}, simultaneous-direction method of multipliers \citep[SDMM; e.g.][]{carrilloPURIFYNewApproach2013}, alternating direction method of multipliers \citep[ADMM; e.g.][]{pratleyRobustSparseImage2018,pratleySparseImageReconstruction2021}, and the proximal gradient method  \citep[][]{caiUncertaintyQuantificationRadio2018a,caiUncertaintyQuantificationRadio2018,caiOnlineRadioInterferometric2019}. Besides reconstruction, proximal methods have also been developed to perform uncertainty quantification for radio interferometric imaging \citep[][]{caiUncertaintyQuantificationRadio2018a,caiUncertaintyQuantificationRadio2018,caiProximalNestedSampling2021}.

In this paper we compare our new learned methods with a primal dual hybrid gradient \citep[PDHG;][]{chambolleFirstOrderPrimalDualAlgorithm2011} method for finding a solution to the constrained analysis problem (Equation~\ref{eq:constrained-analysis}) as described in \citet{onoseScalableSplittingAlgorithms2016}. For the dictionary representation of our signal we chose the SARA representation, i.e.\ a combination of the Dirac basis as well as the first eight Daubechies wavelets, Db1-Db8 \citep[][]{daubechiesTenLecturesWavelets1992}, which was shown to work well for astronomical reconstruction \citep[][]{carrilloSparsityAveragingReweighted2012,carrilloSparsityAveragingCompressive2013}.

\subsection{Learned methods}
Traditional approaches handle the ill-posedness of the inverse problems by using prior information through the use of handcrafted regularizers that, while general, fail to capture detailed prior information of real data.  Learned methods attempt to overcome this by instead enforcing the prior information implicitly specified by training data, ensuring that the prior information promotes images which are in some sense similar to the training data.

Since learned methods rely on learning the prior information for the imaging problem from the data they are provided, reconstructive power depends on the quality and quantity of the training data provided. Furthermore, learned methods have to be trained before evaluation can take place, a process that may take some time, yet only has to be performed once.  Once training is done, imaging can then be performed rapidly.
If training data in the form of input-output pairs are available, the network can be trained in a supervised approach, with the network receiving the input measurements and their respective targets. When input-output pairs are not available, the networks can be learned adversarially, where the network is trained using a distribution of measurements to fit an independent distribution of target outputs, which is sometimes interpreted as semi-supervised learning. If no training outputs are available the methods need to be learned in a self-supervised way.

Learned methods that are completely independent of the measurement operator and learn a direct mapping from measurements to target are generally not effective since they are difficult to train and require huge volumes of training data \citep{adlerSolvingIllposedInverse2017}. Therefore, most learned approaches incorporate the measurement operator in some capacity to encode or leverage the mapping from the measurement to the reconstruction space. The learned network is then constructed around that.  As briefly overviewed in Section~\ref{sec:introduction}, learned data-driven imaging approaches broadly fall into three categories, differentiated by the degree to which model-based information such as the measurement operator is encoded or leveraged.  By making greater use of the measurement operator superior performance can often be achieved but at the cost of increased computational time.  We subsequently review each of these three classes of approach.

\subsubsection{Learned regularization}\label{sec:regularization}
One method for learning the prior information from training data is by using a learned regularizer in conjunction with traditional optimization schemes. These algorithms aim to replace the regularizer $\mathcal{S}$ in Equation~\ref{eq:variational-regularisation} with a learned regularizer, encoding the prior information in the training data.

A common approach is to replace the application of the proximal operators with a learned map to compute the update. These methods typically adopt plug-and-play (PnP) denoisers that are used instead of a proximal update step \citep[][]{venkatakrishnanPlugandPlayPriorsModel2013,ryuPlugandPlayMethodsProvably2019}. A variant of a PnP denoiser was applied to radio interferometry by \citet{terrisImageReconstructionAlgorithms2022}.

Various other pioneering approaches to learned regularization have also been considered.  The regularizer can be formed as the norm of a learned dictionary that translates to a sparse representation of the data \citep{xuLowdoseXrayCT2012} or by using a constraint based on a learned scattering transform \citep[][]{dokmanicInverseProblemsInvariant2016} in place of a traditional regularizer.
Alternatively, methods are used that implement deep neural networks to act as regularizers \citep[e.g.][]{liNETTSolvingInverse2020,koblerTotalDeepVariation2020a} as well as methods that train adversarially learned neural networks \citep[e.g.][]{lunzAdversarialRegularizersInverse2018,mukherjeeLearnedConvexRegularizers2020}.

While learned regularization approaches often achieve excellent reconstruction quality since they make full use of the measurement operator, along with learned prior information, they are computationally demanding since the full measurement operated must be evaluated for each iteration.

\subsubsection{Learned sequential methods} \label{sec:sequential}
An alternative approach is to consider a model that is split up into a sequence of operations. Sequential models are a composition of a learned operator acting in the data space $\boldsymbol{C}_\theta: Y \rightarrow Y$, an adjoint or pseudo-inverse mapping from the data to the reconstruction space, $\boldsymbol{A}:  Y \rightarrow X$, and a learned operator acting in the image space $\boldsymbol{B}_\theta: X \rightarrow X$, where learned operators depend on the parameters $\theta$ \citep[e.g.][]{zhengDualdomainDeepLearningbased2020}:
\begin{equation}\label{eq:sequential}
	\boldsymbol{\Phi}^\dagger_{\theta} = \boldsymbol{B}_{\theta} \circ \boldsymbol{A} \circ \boldsymbol{C}_{\theta}.
\end{equation}

When either the operator in the reconstruction domain, $\boldsymbol{B}_{\theta}$, or in the data domain, $\boldsymbol{C}_\theta$, is set to be the identity operator, we recover learned pre-processing or learned post-processing methods respectively.
Most sequential methods are of the post-processing type since they are relatively easy to train as learning of the network is decoupled from the mapping from measurement to reconstruction space; the post-processing network can thus be trained independently, reducing training time substantially \citep[e.g.][]{jinDeepConvolutionalNeural2017,chenLowDoseCTResidual2017,yiSharpnessAwareLowDoseCT2018}.
Post-processing methods have been used in astronomy using super resolution networks \citep[][]{allamjrRadioInterferometricImage2016, connorDeepRadiointerferometricImaging2022}, denoisers \citep{terrisDeepPostProcessingSparse2019}, and convolutional autoencoders \citep[][]{ghellerConvolutionalDeepDenoising2021}.

Post-processing methods only apply the adjoint or pseudo-inverse of the measurement operator once and so are highly computationally efficient.  However, since the measurement operator is not continually leveraged, reconstruction quality suffers as a consequence.  To compensate sequential models typically employ more elaborate network architectures \citep[such as U-Nets;][]{ronnebergerUNetConvolutionalNetworks2015} than those adopted in learned iterative methods. Nevertheless, learned post-processing approaches can sometimes struggle to achieve the reconstruction quality of learned iterative approaches.

\subsubsection{Learned iterative methods}\label{sec:learned-iterative}
\noindent
Learned iterative methods unroll a small, fixed number of iterations of an iterative solver, e.g. proximal gradient method, and replace the proximal operators with learned convolutional neural networks (CNNs) \citep[][]{gregorLearningFastApproximations2010}. The CNNs used are typically small feed-forward networks with just a few convolutional layers \citep[e.g.][]{yangADMMNetDeepLearning2017,putzkyRecurrentInferenceMachines2017,adlerSolvingIllposedInverse2017}, in contrast to the architectures used in sequential models.  To turn the proximal gradient method of Equation~\ref{eq:forward-backward} into a learned method we replace the proximal operator with a learned network $\boldsymbol{\Lambda}_{\theta,i}$:
\begin{equation}\label{eq:learned2-forward-backward}
	\boldsymbol{x}_{i+1} = \boldsymbol{\Lambda}_{\theta, i}\left(\boldsymbol{x}_{i} - \nabla \mathcal{L} (\Phi \boldsymbol{x}_{i} ) \right),
\end{equation}
where $\boldsymbol{\Lambda}_{\theta,i}$ can be a different network for each iteration $k$ or one network, $\boldsymbol{\Lambda}_{\theta,i} = \boldsymbol{\Lambda}_{\theta}$. The latter scenario results in learning an approximation to the proximal operator.

\citet{hauptmannMultiScaleLearnedIterative2020} propose evaluating the operator at different, increasingly finer resolutions/scales such that the full resolution operator only needs to be evaluated once. Combining this with a multiscale neural net like the U-Net \citep{ronnebergerUNetConvolutionalNetworks2015} cuts down evaluation time (and thus training time) significantly, while also reducing memory requirements.
Other advancements are considered by learning adversarially trained versions of existing models \citep{mukherjeeAdversariallyLearnedIterative2021}. Also, combining approaches using learned regularizers with learned iterative methods, provides algorithms that are closer to the traditional optimization methods and allow to prove rigorous convergence results \citep{mukherjeeEndtoendReconstructionMeets2021}.
Finally, the inclusion of the explicit measurement operator in the learned iterative models improves the robustness and generalizability \citep{boinkPartiallyLearnedAlgorithmJoint2020}, while at the same time reducing the number of trainable parameters and hence reducing the requirement of large training data volumes.

Since learned iterative methods only unroll a small number of iterations, the measurement operator is only evaluated sparingly, making the learned iterative approaches typically faster than the iterative solvers on which they are based.  While the integration of the measurement operator in the network does require its evaluation during training, resulting in considerably longer training times than for the sequential models, encoding the measurement operator in the model typically results in superior performance to learned sequential methods, for only a modest increase in computational time.
\section{Modelling the SPIDER Instrument}\label{sec:modelling-spider}
In order to reconstruct an image from interferometric measurements we need a model for our measurement process in the form of a measurement operator. The accuracy of the measurement operator is a critical factor for the accuracy of the reconstruction and its computational complexity is reflected in both training and reconstruction time.

For interferometric imaging, the measurement process is given by Equation~\ref{eq:visibility}. In a discretized version of Equation~\ref{eq:visibility}, following the notation of Equation~\ref{eq:inverse-problem}, the measurement operator $\boldsymbol{\Phi}: \mathbb{R}^{N}  \rightarrow \mathbb{C}^{M}$ is a non-uniform discrete Fourier transform (NUDFT) mapping an image in $\mathbb{R}^{N}$ to $M$ non-uniformly distributed Fourier measurements.

For an image $\boldsymbol{x} = f(\boldsymbol{\chi})$, with spatial pixel coordinates $\boldsymbol{\chi}_{ij}$  at the $i$-th and $j$-th pixels of the image, and Fourier measurements $\boldsymbol{y} = \hat{f}(\boldsymbol{\xi})$ at non-uniformly distributed Fourier coordinates $\boldsymbol{\xi}_k = (u_k, v_k)$ (in contrast to Equations~\ref{eq:DFT-forward}~and~\ref{eq:DFT-inverse}), the unitary NUDFT can be expressed as
\begin{equation}\label{eq:NUDFT-forward}
    \hat{f}(\boldsymbol{\xi}_k) = \frac{1}{\sqrt{N}} \sum_{{i}=1}^{N_1} \sum_{{j} =1}^{N_2} f(\boldsymbol{\chi}_{{ij}}) \  \mathrm{e}^{-\mathrm{i\,} \boldsymbol{\chi}_{{ij}} \cdot \boldsymbol{\xi}_k},
\end{equation}
which is a finite sum over all pixels $(i,j)$ of an $N_1\times N_2$ image with a total number of $N=N_1\times N_2$ pixels.

To construct the dirty image $f_D(\boldsymbol{\chi})$, the adjoint of the measurement operator is applied, which maps from the non-uniformly distributed Fourier measurements to a uniformly sampled image
\begin{equation}\label{eq:NUDFT-adjoint}
    f_D(\boldsymbol{\chi}_{ij}) =  \frac{1}{\sqrt{N}}\sum_{k=1}^{M}  \hat{f}(\boldsymbol{\xi}_k) \ \mathrm{e}^{\mathrm{i\,} \boldsymbol{\chi}_{ij} \cdot \boldsymbol{\xi}_{k}}.
\end{equation}

We can approximate the inverse Fourier transform (Equation~\ref{eq:FT-inverse}) by evaluating a weighted, finite Fourier series at the non-uniformly distributed Fourier coordinates. The accuracy of the resulting pseudo-inverse DFT is determined by the choice of sampling $\boldsymbol{\xi}_k$ and the corresponding measurement weights $w(\boldsymbol{\xi}_k)$ and is given by
\begin{equation}\label{eq:NUDFT-pseudo-inverse}
    f(\boldsymbol{\chi}_{ij}) \approx  \sum_{k=1}^{M}  w(\boldsymbol{\xi}_k) \hat{f}(\boldsymbol{\xi}_k) \ \mathrm{e}^{\mathrm{i\,} \boldsymbol{\chi}_{ij} \cdot \boldsymbol{\xi}_{k}}.
\end{equation}

The complexity of a naive realization of Equations~\ref{eq:NUDFT-forward},~\ref{eq:NUDFT-adjoint} and \ref{eq:NUDFT-pseudo-inverse} would be $\mathcal{O}(NM)$, however efficient numerical NUDFT schemes make use of interpolation to map the non-uniformly distributed measurements to a uniformly sampled grid and take advantage of the Fast Fourier transform (FFT) to reduce complexity.

In this section we introduce two efficient methods for modelling the measurement operator of the SPIDER instrument. First, we consider the non-uniform Fast Fourier transform \citep[NUFFT;][]{duijndamNonuniformFastFourier1997}, which can be used for arbitrary sampling distributions. Second, we present a new approach that exploits the specific sampling distribution of the SPIDER instrument, making use of similarities of the problem to that of parallel Radon transform of X-ray tomography.

\subsection{NUFFT}\label{sec:nufft}
Using FFTs with a (de)gridding interpolation operator to approximate the non-uniformly distributed Fourier measurements lowers the computational cost significantly. The standard operator used in radio interferometry can be expressed as a composition of operators \citep[e.g.][]{pratleyRobustSparseImage2018}
\begin{equation}\label{eq:forward}
    \boldsymbol{\Phi } =   \boldsymbol{G} \boldsymbol{F} \boldsymbol{Z} \boldsymbol{D},
\end{equation}
where $\boldsymbol{D}: \mathbb{R}^N \rightarrow \mathbb{R}^N$ is an operator that corrects for the gridding operation (discussed further below), $\boldsymbol{Z}: \mathbb{R}^{N} \rightarrow \mathbb{R}^{\alpha^2 N}$ is a zero-padding operator that zero pads the image in the spatial dimension to provide oversampling of a factor $\alpha$ in each direction in the Fourier domain, $\boldsymbol{F}: \mathbb{C}^{\alpha^2 N} \rightarrow  \mathbb{C}^{\alpha^2 N}$ is a 2-dimensional unitary FFT operator, and $\boldsymbol{G}: \mathbb{C}^{\alpha^2 N} \rightarrow \mathbb{C}^{M} $ is the degridding operator that interpolates the measurements off of the uniform Fourier grid via convolution with a spreading kernel centred at the irregularly spaced measurements points in the Fourier domain. The operator $\boldsymbol{D}$ corrects for the convolution with this spreading kernel by a pointwise division of the image with the Fourier transform of the spreading kernel.

Equation~\ref{eq:forward} describes the measurement process, going from the observable  to the measurement space. The adjoint of this operation, $\boldsymbol{\Phi}^*: \mathbb{C}^{M} \rightarrow \mathbb{R}^{N}$ is readily obtained as
\begin{equation}\label{eq:adjoint}
    \boldsymbol{\Phi }^* =  \boldsymbol{D} \boldsymbol{Z}^* \boldsymbol{F}^\dagger \boldsymbol{G}^*,
\end{equation}
where we made use of the self-adjointness of the convolution correction operator($\boldsymbol{D}^* = \boldsymbol{D}$). The gridding operation $\boldsymbol{G}^*: \mathbb{C}^M \rightarrow \mathbb{C}^{\alpha^2N}$  convolves the non-uniformly distributed measurements with the spreading kernel to grid them on a uniformly sampled mesh, on which the inverse FFT, $\boldsymbol{F}^\dagger: \mathbb{C}^{\alpha^2 N} \rightarrow \mathbb{C}^{\alpha^2 N}$, is applied to obtain an image, which is restricted to $\mathbb{R}^N$ via symmetric cropping through the operator $\boldsymbol{Z}^*:\mathbb{R}^{\alpha^2 N} \rightarrow \mathbb{R}^{ N}$, and deconvolved via pointwise division with the Fourier inverse of the spreading kernel, using the operator~$\boldsymbol{D}$, to correct for the gridding operation.

The efficiency of the NUFFT schemes hinges upon a suitable choice of the spreading kernel, in particular its localization in the spatial and frequency domains and the ease of computing the Fourier transform of the kernel for deconvolution. The kernel needs to have a small support in frequency space to be computationally inexpensive, while also having a small support in the image domain as to minimize the effects of aliasing.
Here we adopt the Kaiser-Bessel (KB) kernel \citep{jacksonSelectionConvolutionFunction1991,fesslerNonuniformFastFourier2003}. The kernel is truncated in the Fourier domain to limit its support and therefore its computational cost. Since these kernels are linearly separable, i.e.\ the kernel can be written $\hat{c}(u, v) = \hat{c}(u)\hat{c}(v)$, we consider the kernel in only one dimension:
\begin{equation}
    \hat{c}(u) = \begin{cases}
        \frac{1}{J} \frac{I_0\left(\beta \sqrt{1-(\frac{2u}{J})^2}\right)}{I_0(\beta)} & \quad \text{for } |u| \leq J/2, \\
        0                                                                              & \quad \text{for } |u| > J/2,
    \end{cases}
\end{equation}
where $I_0$ is the zeroth-order modified Kaiser-Bessel function, $\beta$ determines the shape of the kernel, and $J$ the support of the kernel.
The correction for applying this convolutional kernel can be calculated analytically by taking the inverse Fourier transform of the kernel to get \citep{jacksonSelectionConvolutionFunction1991,fesslerNonuniformFastFourier2003}
\begin{equation}
    c(x) = \left[ \frac{\sin \left( \sqrt{\pi^2 x^2 J^2 - \beta^2}\right)}{\sqrt{\pi^2 x^2 J^2 - \beta^2}}\right]^{-1}.
\end{equation}
The correction operator can then be found by calculating
\begin{equation}
    D(\boldsymbol{\chi}_{ij}) = c\left(\frac{i}{N_1} -\frac{1}{2}\right) \ c\left(\frac{j}{N_2} -\frac{1}{2}\right).
\end{equation}

Using $\beta = 2.34 J$ for the spread of this Kaiser-Bessel kernel gives similar performance to the optimal min-max interpolation kernel proposed by \citet{fesslerNonuniformFastFourier2003}.

The pseudo-inverse of the NUFFT is obtained by including the measurement weights in the Fourier sum
\begin{equation}\label{eq:pseudo-inverse}
    \boldsymbol{\Phi}^\dagger = \boldsymbol{\Phi}^* \boldsymbol{W} = \boldsymbol{D} \boldsymbol{Z}^* \boldsymbol{F} \boldsymbol{G}^* \boldsymbol{W},
\end{equation}
where $\boldsymbol{W}$ is a diagonal matrix with elements $\boldsymbol{W}_{kk} = w(\boldsymbol{\xi}_k)$ given by the weight for each of the measurements.

In radio astronomy, interferometric measurements are typically weighted according to one of three weighting schemes: natural, uniform, or robust weighting \citep{taylorSynthesisImagingRadio1999}. Natural weighting maximizes the sensitivity of the observation and weights each visibility by the uncertainty on the measurement $\boldsymbol{W}^{\text{Natural}}_{k,k} = \sigma_k^{-2}$, with $\sigma_k$ the uncertainty on the measured visibility $\boldsymbol{y}_k$. Natural weighting is akin to statistical whitening of data and ensures unit variance of the uncertainty on the measurements and gives the highest signal-to-noise ratio for detecting weak sources.

Since there is typically a higher density of samples near the $(u,v)$ origin in radio interferometry, natural weighting emphasizes low-frequency measurements, which can be undesirable when imaging sources with both large- and small-scale structure. Uniform weighting is an alternative weighting scheme that accounts for this by weighting the density of the sampling distribution by $\boldsymbol{W}^{\text{Uniform}}_{k,k} = \sigma_k^{-2} N_s^{-1}(k)$, where $N_s(k)$ is typically the number of measurements in a symmetrical region (either square or circular) with width $s$ around the measurement $\boldsymbol{y}_k$. If the measurement uncertainties are varying this can be replaced by the sum of the reliability weights in this region $N_s = \sum_{k'} \sigma^{-2}_{k'}$, where $k'$ are the indices of the measurements in the local region around measurement $k$. The width $s$ of the region is typically chosen to be the size of the size of the elements of the Fourier grid.

Lastly, robust weighting considers a trade-off between natural and uniform weighting, controlled by a robustness parameter. In this paper we use uniform weighting and weight the measurements according to the density of the sampling distribution as well as their uncertainties.

\subsection{Sub-scale operators}\label{sec:sub-scale-operators}
An approach to reduce the computational cost of the measurement operator is to evaluate it at a lower resolution and restricted Fourier space. The forward and adjoint of the measurement operator can be evaluated at a series of scales by applying a filter bank of low pass and high pass partition of unity filters; see \citet{panLearningInvisiblePhotoacoustic2022} for the details of such filter banks. Instead of using smooth partition of unity filters as used in \citet{candesFastDiscreteCurvelet2006,panLearningInvisiblePhotoacoustic2022}, we use binary 0-1 partition of unity filters. Since the NUFFT evaluates the FFT on an upsampled grid  by zero-padding the image, the periodicity in the image domain and the corresponding decay of the Fourier coefficients are reinstated. We refer to these sub-scale measurement operators as $\boldsymbol{\Phi}_i: \mathbb{R}^{N_i} \rightarrow \mathbb{C}^{M_i}$, its adjoint $\boldsymbol{\Phi}_i^*: \mathbb{C}^{M_i} \rightarrow \mathbb{R}^{N_i}$, and its pseudo-inverse $\boldsymbol{\Phi}_i^\dagger: \mathbb{C}^{M_i} \rightarrow \mathbb{R}^{N_i}$, operating on the reduced image scale $N_i$ and the number of measurements restricted to the reduced Fourier space, $M_i$. Since these sub-scale operators are evaluated on a smaller image scale as well as a restricted Fourier space, they are considerably more computationally efficient than the full-scale measurement operators.

\subsection{NU-Radon method}\label{sec:nu-radon}
For the NUFFT the measurement weights are determined by considering the 2D sampling density of the instrument. In this section we propose an alternate, principled approach to derive these weights based on the similarities of the SPIDER sampling to the sampling induced by the 2D Radon transform.

The Radon transform,
\begin{equation}\label{eq:radon-transform}
    \boldsymbol{R}f(r, \phi) = \boldsymbol{R}_\phi f(r) = \int_{ x \cdot \phi = r} f(x) \mathrm{d}x,
\end{equation}
models the linear attenuation of photons passing through an object. It maps the attenuation to a family of integrals along parallel lines, parametrized in polar coordinates $\phi$, the angle of the projection, and $r$ the radial distance from the origin along that line. The Fourier slice theorem
\begin{equation}\label{eq:fourier-slice-theorem}
    \hat{\boldsymbol{R}_\phi f} (\rho) = \hat{f}(\rho, \phi),
\end{equation}
expresses the equivalence of the 2D Fourier transform in polar coordinates and the 1D Fourier transform along the detector coordinate of the Radon transform, where $\rho$ is the distance in the Fourier domain along the radial spoke at angle $\phi$.

The adjoint operation of the Radon transform is the back-projection which can be used to recover the dirty reconstruction through
\begin{equation}\label{eq:back-projection}
    f_D(r, \phi) = \int_{0}^{\pi} \int_{-\infty}^{\infty} \hat{\boldsymbol{R}_\phi f}(\rho) \ \mathrm{e}^{\mathrm{i\,} 2 \pi r \rho }  \ \mathrm{d}\rho \ \mathrm{d}\phi.
\end{equation}
The inverse of the radon transform is the \emph{filtered} back-projection and includes a ramp filter $|\rho|$, which corresponds to the Jacobian of the change from Cartesian to polar coordinates and which boosts the power of high-frequency measurements:
\begin{equation}\label{eq:filtered-back-projection}
    f(r, \phi) = \int_{0}^{\pi} \int_{-\infty}^{\infty} \hat{\boldsymbol{R}_\phi f}(\rho) \ \mathrm{e}^{\mathrm{i\,} 2 \pi r \rho } \ |\rho| \ \mathrm{d}\rho \ \mathrm{d}\phi .
\end{equation}

Using the Fourier slice theorem we can express the NUDFT (Equation~\ref{eq:NUDFT-forward}) in terms of the Radon transform, and in polar coordinates, by
\begin{equation}
    \hat{f}(\boldsymbol{\xi}_k) = \hat{f}(\rho_k, \phi_k) = \frac{1}{\sqrt[4]{N}} \sum_{i=1}^{\sqrt{N}} \boldsymbol{R}_{\phi_k} f(r_i) \ \mathrm{e}^{-\mathrm{i\,} r_i \rho_k},
\end{equation}
where the measurements are gathered by taking a 1D NUDFT of the Radon transform of the signal along a spoke at angle $\phi_k$ and $\sqrt{N}$ corresponds to the diameter of the circle inscribed into the square image. Similarly, we discretize the back-projection (Equation~\ref{eq:back-projection}) to obtain the dirty reconstruction 
\begin{equation}
    f_D(\boldsymbol{\chi}) = \frac{1}{\sqrt[4]{N}}\sum_{p = 1}^{P}\sum_{q=1}^{Q} \hat{\boldsymbol{R}_{\phi_p}f}(\rho_q) \ \mathrm{e}^{\mathrm{i\,} \rho_q \boldsymbol{\chi} \cdot \phi_p},
\end{equation}
where we Fourier transform the $Q$ non-uniformly distributed Fourier measurements at radial distances $\rho_q$ for each of the spokes, followed by back-projecting the signals for each of the $P$ projection angles $\phi_p$.

Both the forward and adjoint operations of the measurement operator use a 1D NUDFT which can be accelerated by using (de)gridding in the form of a 1D NUFFT. The combined method of using the Radon transform and the NUFFT can be described as:
\begin{equation}
    \boldsymbol{\Phi} = \boldsymbol{N} \boldsymbol{R},
\end{equation}
where $\boldsymbol{{R}}: \mathbb{R}^N \rightarrow \mathbb{R}^{P\sqrt{N}}$ is the Radon transform, and $\boldsymbol{N}: \mathbb{C}^{M} \rightarrow \mathbb{R}^{P\sqrt{N}}$ is the 1D NUFFT. Similarly, the adjoint operation is given by
\begin{equation}
    \boldsymbol{\Phi}^* = \boldsymbol{R}^* \boldsymbol{N}^*,
\end{equation}
where $\boldsymbol{{R}}^*:\mathbb{R}^{P\sqrt{N}}  \rightarrow \mathbb{R}^N$ is the back-projection operation, and $\boldsymbol{N}^*: \mathbb{C}^{M} \rightarrow \mathbb{C}^{N} $ is the 1D adjoint NUFFT.

To obtain the pseudo-inverse of the measurement operator, part of the measurement weights are now informed by the ramp filter in the inverse of the Radon transform (Equation~\ref{eq:filtered-back-projection}). The ramp filter for our non-uniformly distributed measurements are calculated as $w_{\text{Radon}}(\boldsymbol{\xi}) = \|\boldsymbol{\xi}\|_{\ell_2}$. The pseudo-inverse of the measurement operator is then defined as
\begin{equation}\label{nuradon-inverse}
    \boldsymbol{\Phi}^\dagger = \boldsymbol{R}^\dagger \boldsymbol{N}^\dagger,
\end{equation}
where $\boldsymbol{{R}}^\dagger:\mathbb{R}^{P\sqrt{N}}  \rightarrow \mathbb{R}^N$ is the filtered back-projection (with weights $w_\text{radon}(\boldsymbol{\xi})$), and $\boldsymbol{N}^\dagger: \mathbb{C}^{M} \rightarrow \mathbb{C}^{N} $ is the 1D inverse NUFFT with measurements weights based on the sampling density of the lenslets along one spoke.

\subsection{Comparison}
We implement both the NUFFT and NU-Radon methods in our \texttt{LeIA}\footnote{\url{https://github.com/astro-informatics/LeIA}}code. The NUFFT approach is implemented on the CPU using \texttt{Numpy}\footnote{\url{https://numpy.org/}} as a backend, as well as with GPU acceleration implemented in \texttt{TensorFlow}\footnote{\url{https://www.tensorflow.org/}}. The NU-Radon method is implemented by using a modified version of the \texttt{SciKit-Image}\footnote{\url{https://scikit-image.org/}} Radon transform.

An overview of the computational complexity of the algorithms can be found in Table~\ref{tab:complexity}, where the complexity is split into three operations: calculating the Fourier transform, (de)gridding non-uniformly distributed measurements, and computing the Radon transform (when relevant)\footnote{Recall $N$ denotes the number of image pixels, $M$ the number of measurements, $J$ the support of the spreading kernel, and $P$ the number of spokes of the SPIDER instrument.}. The NUDFT is computationally the most expensive approach, since the evaluation of the discrete Fourier transform takes $\mathcal{O}(NM)$ operations. Both the NUFFT and NU-Radon reduce the computational complexity by utilizing the FFT for calculating the Fourier transform, though these gains are offset by the computational cost for the gridding of the measurements and/or the calculation of the (inverse) Radon transform.

For the particular SPIDER configuration considered ($N=256\times 256$, $M=4440$, and $P=37$), the NUFFT is the best approach in terms of computational complexity. It might be beneficial to use the NU-Radon method when the number of measurements is large (and therefore the computation time is dominated by the (de)gridding operations) or when the number of image pixels is large and the number of spokes is modest ($P\leq \log{N}$).

\begin{table*}
    \caption{The computational complexity of the different methods for modelling the measurement operator of the SPIDER instrument in terms of the number of pixels $N$, the number of measurements $M$, the number of projection angles $P$, and the support of the gridding kernels in pixels $J$.  }
    \label{tab:complexity}
    \begin{tabular}{lllll}
        \hline
                 & Fourier Transform                     & Gridding            & Radon Transform   & Total                                           \\ \hline
        NUDFT    & $\mathcal{O}(NM)$                     & -                   & -                 & $\mathcal{O}(NM)$                               \\
        NUFFT    & $\mathcal{O}(N \log N)$               & $\mathcal{O}(MJ^2)$ & -                 & $\mathcal{O}(N \log N + MJ^2 )$                 \\
        NU-Radon & $\mathcal{O}(\sqrt{N} \log \sqrt{N})$ & $\mathcal{O}(MJ)$   & $\mathcal{O}(NP)$ & $\mathcal{O}(\sqrt{N} \log \sqrt{N} + MJ + NP)$ \\ \hline
    \end{tabular}
\end{table*}

\begin{figure*}
    \centering
    \includegraphics[width=\textwidth, trim= 6cm 1cm 2cm 1cm, clip]{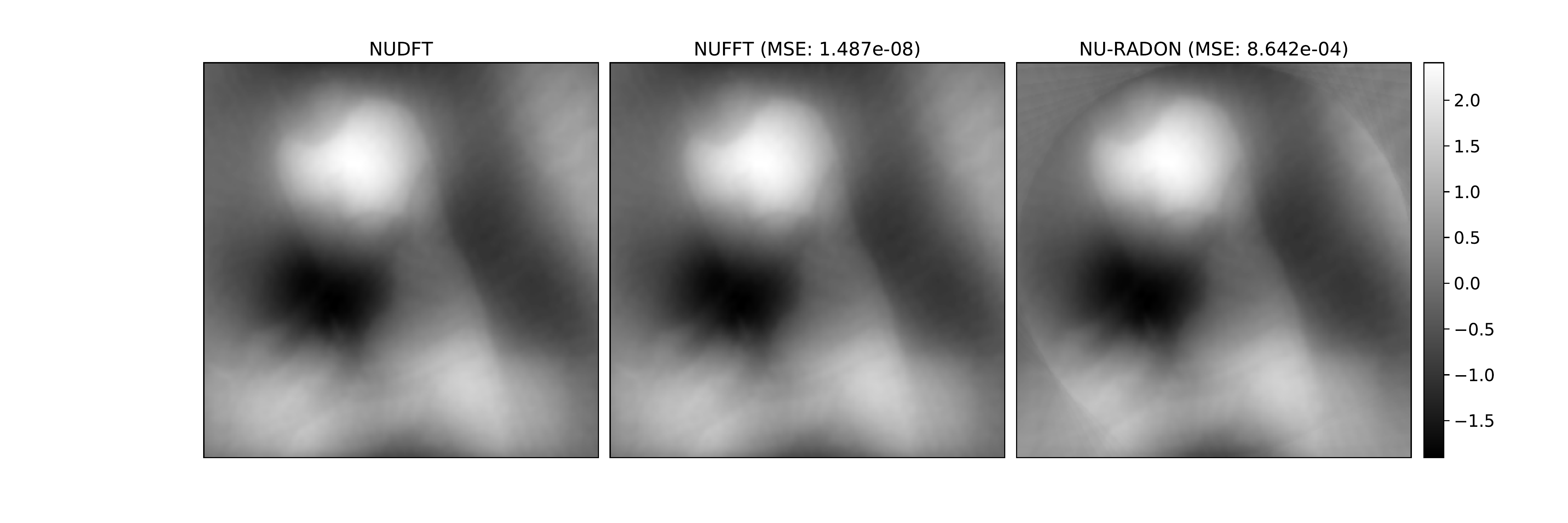}
    \caption{Generating a dirty reconstruction by applying the adjoint of the measurement operator, using the NUDFT, to interferometric measurements of the SPIDER instrument (left), compared to accelerated methods for approximation of the adjoint operation using the NUFFT (centre) and the NU-Radon (right) approaches as detailed in sections \ref{sec:nufft} and \ref{sec:nu-radon} respectively. The MSE of the two accelerated approaches compared to the NUDFT is shown, where the error is only calculated inside the circular aperture that limits the NU-Radon method.}
    \label{fig:m51-self-adjoint}
\end{figure*}

Figure~\ref{fig:m51-self-adjoint} shows the effect of reconstructing a dirty image using the adjoint of the measurement operator applied to a set of simulated Fourier measurements taken with the SPIDER instrument. For the configuration of the SPIDER instrument, it appears that the NUFFT is a more accurate approximation to the NUDFT.

Figure~\ref{fig:m51-self-adjoint} indicates that the dirty reconstructions are mostly dominated by low-frequency structures. The measurement weights used in the pseudo-inverse operations boost the power of high-frequency measurements by weighting the measurements based on the 2D sampling density for the NUFFT and the 1D sampling density along each spoke for the NU-Radon approach (which also includes the weights of the filtered back-projection).

\begin{figure*}
    \centering
    \includegraphics[width=\textwidth, trim= 7.5cm 1.5cm 5.5cm 0cm, clip]{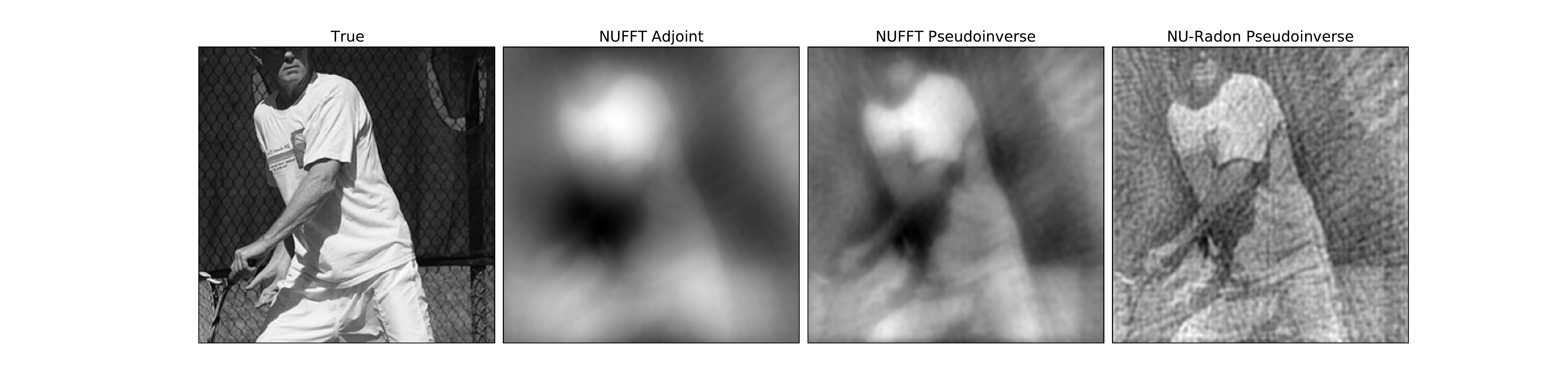}
    \caption{ Measurements are simulated from the original image from the COCO dataset of natural images shown in the left-most image, using a NUFFT with the SPIDER sampling pattern. Generating images by applying adjoint NUFFT operation, the pseudo-inverse NUFFT operation, and the pseudo-inverse NU-Radon operation are shown respectively from left to right.}
    \label{fig:m51-pseudo-inverse}
\end{figure*}

Figure~\ref{fig:m51-pseudo-inverse} compares the different pseudo-inverse implementations. The figure indicates that the pseudo-inverse NUFFT provides a good trade-off between introducing high-frequency information and not having strong high-frequency artefacts. The pseudo-inverse NU-Radon implementation provides sharper features, yet also includes more high-frequency artefacts. In the remainder of the paper we use the NUFFT approach for our experiments. Due to the lower computational cost of its implementation and its generalizability to any sampling distribution for future applications.

\section{Learned Interferometric Imaging for SPIDER}\label{sec:solving-approaches}

In this section we present our two learned approaches for the \mbox{SPIDER} imaging problem.  Both approaches reconstruct images from interferometric measurements and use the adjoint or pseudo-inverse of the measurement operator to create an initial reconstruction. Our approaches fit within the exiting learned post-processing framework \citep[e.g.][]{jinDeepConvolutionalNeural2017}, with the second method introducing a novel post-processing architecture.

The first method is a post-processing learned sequential method to improve the initial reconstruction estimated by the pseudo-inverse and remove artefacts.  This approach is highly computationally efficient, requiring only one application of the pseudo-inverse of the measurement operator to compute the initial reconstruction. We expect final reconstruction quality to be good but limited since no further knowledge of the measurement operator is exploited beyond the initial reconstruction.

The second approach is a learned iterative method and differs to the former by including updates to the reconstruction that utilize knowledge of the measurement operator.  Since additional application of the measurement operator is required this results in a modest increase in computational time compared to our post-processing method.  However, we expect this to be offset by an increase in reconstruction quality due to further exploitation of the measurement operator.

We do not consider learned regularization approaches further since the iterative nature of such approaches would result in a significant increase in computational time, and our primary objective is a very low computational cost during imaging in order to facilitate real-time imaging with SPIDER.

\subsection{Learned post-processing (U-Net)}
\begin{figure}
    \centering
    \includegraphics[width=\columnwidth, trim= 0 1cm 0 0cm, clip]{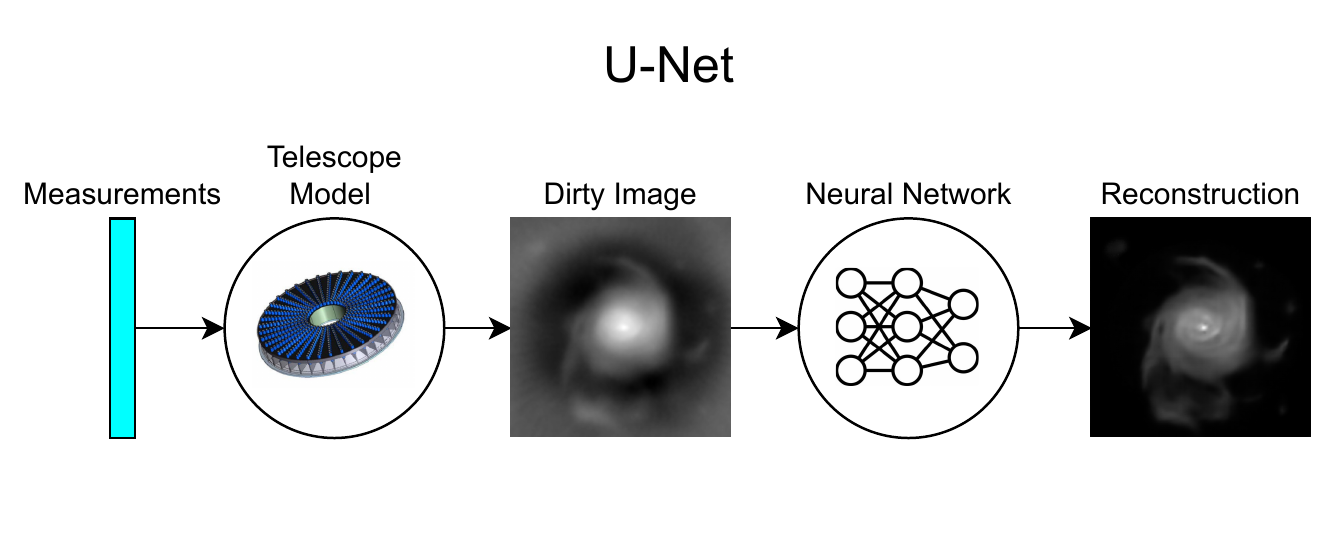}
    \caption{The learned post-processing approach takes the interferometric measurements and uses the telescope model to create an initial reconstruction, in our case the dirty image. This is then passed through the learned post-processing network to create the final reconstruction.}
    \label{fig:u-net-pipeline}
\end{figure}

Learned post-processing methods are a special case of the sequential methods discussed in Section~\ref{sec:sequential}, where the learned operator applied in the data domain, $\boldsymbol{C}_{\theta}$, is replaced with an identity operator. This makes the networks easier to train as the training is performed in the image domain without the need to invoke the measurement operator, speeding up the training process. Our method only post-processes an initial reconstruction and is in this sense similar to a denoiser. The solution can be written as:
\begin{equation}
    \boldsymbol{x}^\star = \boldsymbol{\Phi}^\dagger_\theta \boldsymbol{y} = \boldsymbol{\Lambda}_{\theta}  \boldsymbol{\Phi}^\dagger \boldsymbol{y},
\end{equation}
with $\boldsymbol{\Lambda}_{\theta}$ the learned network, and $\mathcal{\boldsymbol{\Phi}}^\dagger$ the pseudo-inverse of the measurement operator. Figure~\ref{fig:u-net-pipeline} shows a schematic representation of this approach.

The architecture of our network is based on the U-Net \citep[][]{ronnebergerUNetConvolutionalNetworks2015} network architecture, which was originally designed for biomedical image segmentation. U-Nets can also be used for denoising problems by replacing the two segmentation specific output layers with a single output layer returning the denoised image. In this form, U-Net is a convolutional autoencoder with skip connections at each of the scales. Our network includes blocks of 2D convolutions with batch normalization and ReLU activation functions at each scale, followed by MaxPool layers to sub-sample on a downward branch. In the decoder branch, the upsampling is achieved by transposed convolution layers.  A schematic representation of the architecture can be found in Figure~\ref{fig:networks}.

Our learned post-processing is computationally efficient as it only evaluates the measurement operator once per image for both evaluation and training. However, this also limits the extent to which the network can utilize the measurement model.


\subsection{Learned iterative method (GU-Net)}\label{sec:gunet}
\begin{figure}
    \centering
    \includegraphics[width=\columnwidth, trim= 0 .5cm 0 0cm, clip]{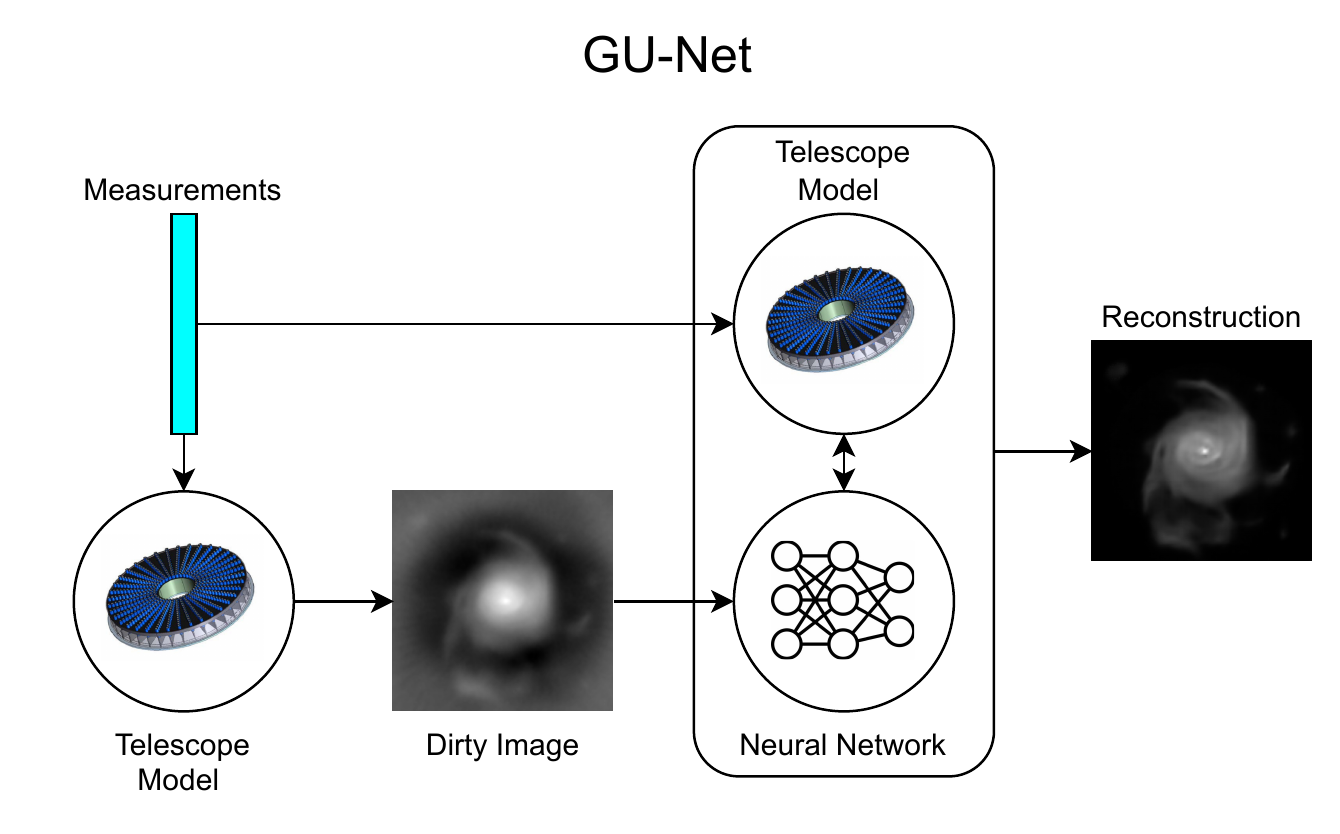}
    \caption{For the learned iterative approach, the interferometric measurements are used to create an initial reconstruction using the telescope model, in our case this results in the dirty image. This is then used as input for the learned iterative network. At several stages of the network, the telescope model is used together with the original measurements to provide model-based updates to the reconstruction.}
    \label{fig:gu-net-pipeline}
\end{figure}

\begin{figure*}
    \centering
    \includegraphics[width=\textwidth]{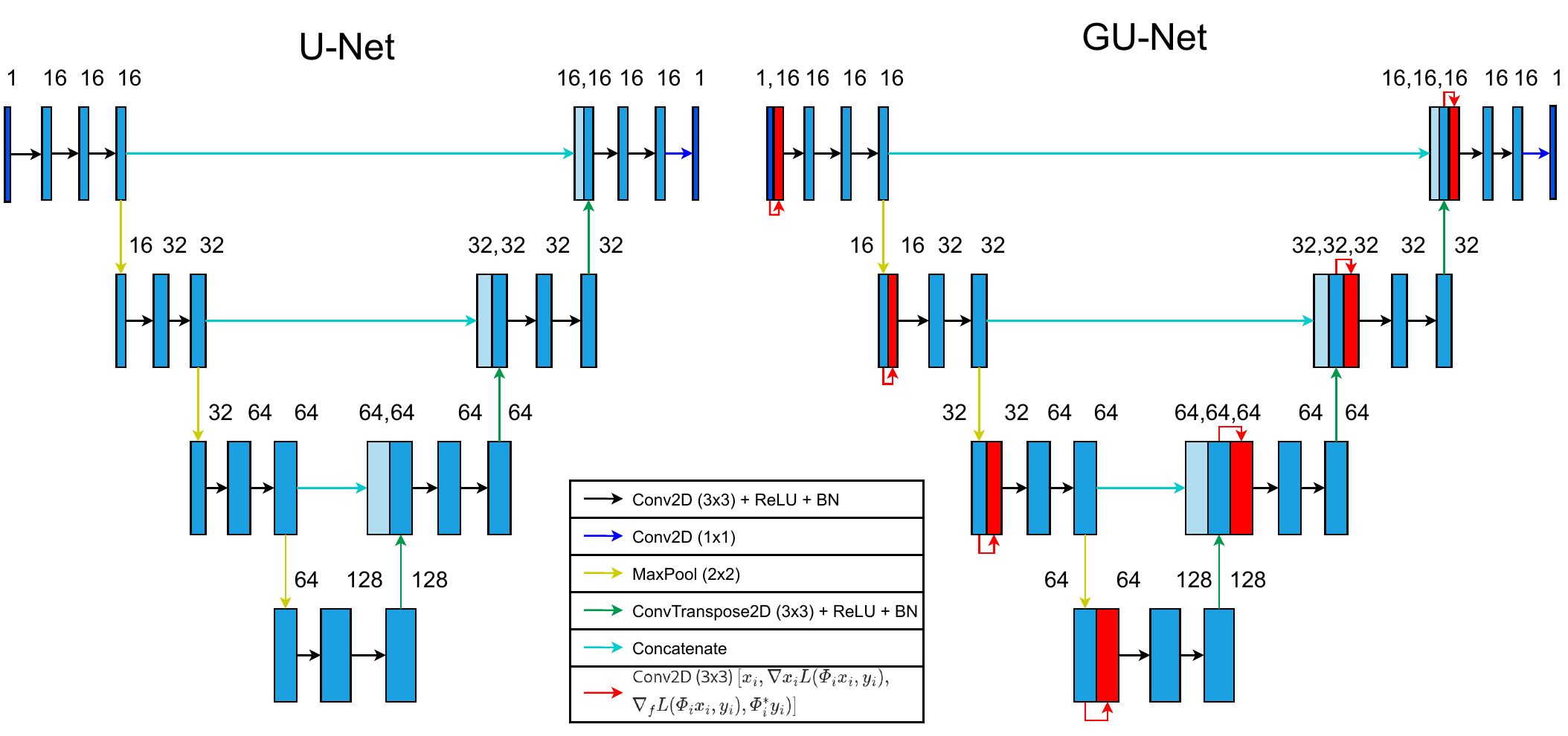}
    \caption{\emph{Left:} The neural network architecture for the U-Net model used for learned post-processing of the initial reconstruction estimated by the pseudo-inverse of the measurement operator applied to the noisy measurements. \emph{Right:} The neural network architecture for the GU-Net model for the learned iterative approach used to reconstruct an image from interferometric measurements. The input image of the network is the pseudo-inverse of the measurement operator applied to the noisy measurements. The model-based operation in this network takes the first channel on each of the scales and calculates the gradient (Equation~\ref{eq:gradient}) and filtered gradient (Equation~\ref{eq:filtered-gradient}) of the data fidelity, as well as a scale-restricted dirty image (Equation~\ref{eq:dirty}). These are combined using a CNN layer (Equation~\ref{eq:gunet-update}), concatenated to the original channels, and passed on through the rest of the network. }
    \label{fig:networks}
\end{figure*}

An alternative approach is to apply learned iterative methods that evaluate the measurement operator several times to improve the reconstruction quality by including measurement information at several stages of the reconstruction process. While some methods in this category closely mimic optimization methods, the method we propose takes inspiration from multiscale methods. Our approach, the Gradient U-Net (GU-Net), expands on the U-Net architecture by calculating the gradient of the data fidelity after every downsampling and after every  upsampling operation, as well as after the input layer, to incorporate measurement information at every scale of the network. Figure~\ref{fig:gu-net-pipeline} shows a diagram summarising the learned iterative reconstruction process. How we implement the network is described below and a schematic representation of the network can be found in Figure~\ref{fig:networks}.

In order to add this measurement information at different scales of the network we use sub-scale operators, $\boldsymbol{\Phi}_i: \mathbb{R}^{N_i} \rightarrow \mathbb{C}^{M_i}$  (see Section~\ref{sec:sub-scale-operators}), where the size of the image space $N_i$ decreases by powers of four (the images are halved in each direction; $N_i = N/4^{i}$), and the size of the measurement space $M_i$ is determined by applying a low-pass filter that restricts the measurements to $\{(u,v):  u \leq u_{\text{max}}/2^{i} \land  v  \leq v{_\text{max}}/2^{i} \}$, where $u_{\text{max}}$ and $v_{\text{max}}$ are the longest baselines observed in each direction.

The first method to incorporate the measurement operator into the reconstruction process is to calculate the gradient of the data fidelity at each scale of the network: $\mathcal{L}( \boldsymbol{\Phi}_i\boldsymbol{x}_i, \boldsymbol{y}_i) = \frac{1}{2} \| \boldsymbol{\Phi}_i\boldsymbol{x}_i - \boldsymbol{y}_i\|_{\ell_2}^2$. This is given by
\begin{equation}\label{eq:gradient}
    \nabla_{\boldsymbol{x}_i} \mathcal{L}( \boldsymbol{\Phi}_i\boldsymbol{x}_i, \boldsymbol{y}_i) \propto  \boldsymbol{\Phi}_i^*(\boldsymbol{\Phi}_i\boldsymbol{x}_i- \boldsymbol{y}_i),
\end{equation}
where $x_i \in \mathbb{R}^{N_i}$ is the first channel of the current iterate at scale $i$ of the network and $y_i \in \mathbb{C}^{M_i}$ is the low-pass filtered measurement vector of visibilities.

By analogy to the Radon transform, the gradient does not contain the high frequency features which are reinstated via the ramp filter in the filtered back-projection. Therefore, we also include the filtered gradient where the high-frequency features are boosted using a sampling density based filter. This is analogous to taking the gradient of a weighted least-squares norm with the measurement weights the same as the uniform weights described in Section~\ref{sec:nufft}:
\begin{equation}\label{eq:filtered-gradient}
    \nabla_{\boldsymbol{x}_i}^{f} \mathcal{L}( \boldsymbol{\Phi}_i\boldsymbol{x}_i, \boldsymbol{y}_i) \propto  \boldsymbol{\Phi}_i^*(\boldsymbol{W}_i^{\text{Uniform}}(\boldsymbol{\Phi}_i\boldsymbol{x}_i- \boldsymbol{y}_i)),
\end{equation}
where $\boldsymbol{W}_i^{\text{Uniform}} \in \mathbb{R}^{M_i \times M_i}$ is a diagonal matrix with the measurement weights for the sub-selected measurements as the diagonal elements.

In addition to the gradient information, we also add the scale-restricted dirty reconstruction (adjoint of the measurement operator applied to the noisy measurements):
\begin{equation}\label{eq:dirty}
    \boldsymbol{x}_{i, \text{dirty}} = \boldsymbol{\Phi}_i^* \boldsymbol{y_i}.
\end{equation}
In doing so, the network can learn an update between the dirty reconstruction, the current iterate passed down by the network, and the gradient and filtered gradient information based on the current iterate.

The measurement information encoded in the (sub-scale) gradient and filtered gradient, and the dirty reconstruction is added at each scale of the U-Net architecture after downsampling and after upsampling. The added measurement information at each scale of the network is given by
\begin{equation}\label{eq:gunet-update}
    \tilde{\boldsymbol{x}}_{i} = \boldsymbol{\Lambda}_{i,\theta}\left(\boldsymbol{x}_i, \ \nabla_{\boldsymbol{x_i}} \mathcal{L}( \boldsymbol{\Phi}_i\boldsymbol{x}_i, \ \boldsymbol{y}_i), \nabla_{\boldsymbol{x_i}}^f \mathcal{L}( \boldsymbol{\Phi}_i\boldsymbol{x}_i, \ \boldsymbol{y}_i) ,\boldsymbol{\Phi}_i^* \boldsymbol{y_i}\right),
\end{equation}
where $\boldsymbol{\Lambda}_{i,\theta}$ is the learned convolutional operator to combine the measurement operator information. At each scale of the network, the gradient information is calculated for the first channel of the current iterate. The learned convolutional operator takes the three channels (gradient, filtered gradient, and dirty reconstruction) representing the measurement information and convolves it with a number of filters matching the number of channels of the current iterate in the network, so as not to dilute the measurement operator information when combining it with the original channels.

Evaluating the added measurement information in Equation~\ref{eq:gunet-update} results in two evaluations of the measurement operator (one each for the gradient and filtered gradient) and three evaluations of the adjoint operator (one each for the gradient and filtered gradient and one for the scale-restricted dirty reconstruction). Since the measurement information is added after down- and upsampling, this results in four full-scale evaluations of the forward and six full-scale evaluations of the adjoint of the measurement operator for the added measurement information. An additional  evaluation of a single full-scale pseudo-inverse of the measurement operator is also required for the initial reconstruction.

This method is computationally more expensive than the U-Net post-processing method because the measurement operator has to be evaluated several times. However, as most of the evaluations of the measurement operator are at reduced scales, the impact on reconstruction time is kept to a minimum. Furthermore, inclusion of multiple iterations with the (albeit downsampled) measurement operator enhances the impact of the model on the reconstruction, improving robustness and generalizability of the model. Similar approaches that leverage multiscale evaluation of the forward/adjoint operators to reduce computational cost with application in X-ray CT imaging were proposed in \citet{hauptmannMultiScaleLearnedIterative2020} and \citet{trentArchitectureInspiredSolvers2020}.


\section{Experiments and results}\label{sec:results}
In order to evaluate the reconstruction performance and computational cost of the learned imaging methods proposed we perform reconstructions on simulated measurements and compare the reconstructions the ground truths. To assess reconstruction quality we compute the peak signal-to-noise ratio (PSNR) to measure how well the reconstruction matches the measurements and the structural similarity index measure \citep[SSIM;][]{wangImageQualityAssessment2004} to assess the structural similarity of the reconstructions.  We compare the two learned approaches described in Section~\ref{sec:solving-approaches} to using a reconstruction obtained with the pseudo-inverse (Equation~\ref{eq:pseudo-inverse}) and a primal-dual method as described in Section~\ref{sec:prox-algorithms} (the implementation uses \texttt{OptimusPrimal}\footnote{\url{https://github.com/astro-informatics/Optimus-Primal}} and runs for 300 iterations), which produces results on par with the current state-of-the-art in astronomical interferometric reconstruction. Specifically, we discuss the different datasets used, the measurement simulation process, network training, and reconstruction results. Besides assessing the computational cost and the reconstruction  quality of the different approaches, we also evaluate the robustness of the methods proposed to additional noise and generalisability to other smaller datasets through transfer learning.

Implementations for the measurement operators, the learned imaging techniques and the routines used to simulate interferometric measurements and train the neural networks can be found in our \texttt{LeIA}\footnote{\url{https://github.com/astro-informatics/LeIA}} codebase.

\subsection{Datasets}\label{sec:datasets}

We consider three datasets with different characteristics to evaluate the reconstruction performance of the imaging methods.
The three datasets cover different potential use cases of the SPIDER instrument, including standard imaging, astronomical imaging and Earth observation.
All images are converted to greyscale and cropped to a size of $256 \times 256$. For each epoch of training the images are rotated and flipped randomly, and measurements are simulated and contaminated with random Gaussian noise as described in Section~\ref{sec:simulating-measurements}. Note that the networks could also be trained at larger image sizes, however since the SPIDER instrument only sparsely samples the Fourier domain (SPIDER acquires at a sparsity of $\frac{4440}{256 \times 256} \approx 7\%$), reconstructing at a larger image size would be challenging.

Since these methods are fully convolutional networks they can be adapted to reconstruct at larger image sizes, if the measurement operator is adapted accordingly. Note however that the SPIDER instrument only sparsely samples the Fourier domain (SPIDER acquires at a sparsity of $\frac{4440}{256 \times 256} \approx 7\%$), because of this sparsity reconstructing at a larger image size will not necessarily yield reconstructions at higher resolution.

The Common Objects in COntext \citep[COCO;][]{linMicrosoftCOCOCommon2014} dataset is a large and diverse set of natural images. We selected a subset of 3000 images split into 2000 training and 1000 test images. This dataset provides a large and diverse dataset of natural images for training the networks. Instruments similar to SPIDER could in future be considered as standard imaging devices for natural images. Furthermore, large datasets of readily available natural images such as this can be used for transfer learning, as considered subsequently. This dataset is initially used to demonstrate the general reconstruction performance of the reconstruction methods when trained on a large set of diverse images. All the images in this dataset have dimensions larger than our network input size, and during training, a random region of $256 \times 256$ is cropped out of the images at every epoch.

Next, we consider a small domain-specific dataset of 450 simulated galaxy images, split into 300 training and 150 test images. The galaxy images are obtained from the IllustrisTNG simulations \citep{nelsonIllustrisTNGSimulationsPublic2019}. The H-alpha column densities from the TNG50 simulation of the IllustrisTNG project \citep{nelsonFirstResultsTNG502019,pillepichFirstResultsTNG502019} are binned to a $256\times 256$ grid to obtain images of simulated galaxy structures. Pixels with no simulation data are inpainted using a primal-dual method in an unconstrained setting using the $\ell_1$-norm of a dictionary of wavelets containing the Dirac basis and the first eight Daubechies wavelets (Db1-Db8) as the regularization functional. We add a small amount of noise to the simulation data ($\text{ISNR}=30\text{dB}$) and use Bayesian inference to estimate the regularization parameter for this inpainting problem from the simulation data as described in \citet{pereyraMaximumaposterioriEstimationUnknown2015}.  These steps are followed merely to process the IllustrisTNG particle simulations in order to provide a dataset of ground-truth galaxy images suitable for our imaging experiments.

The final dataset we consider is another small domain-specific dataset of 450  satellite Earth observations taken from the Deep Globe satellite challenge \citep{demirDeepGlobe2018Challenge2018}, split into 300 training and 150 test images.

\subsection{Simulating measurements}\label{sec:simulating-measurements}

For a SPIDER instrument with parameters as defined in Table~\ref{tab:spider-config}, we simulate measurements using the NUFFT operator (Section~\ref{sec:nufft}) with upsampling factor 2 per dimension, resulting in upsampling to $512 \times 512$ images, and a $6 \times 6$ KB kernel, which in this configuration results in $y \in \mathbb{C}^{4440}$ measurements.

We contaminate the measurements with complex Gaussian noise, $\Re (\boldsymbol{n}), \Im (\boldsymbol{n}) \in \mathcal{N}(0, \sigma /\sqrt{2})$ with 0 mean and standard deviation of the real and imaginary components defined by an input signal-to-noise ratio (ISNR) of $30\text{dB}$
\begin{equation}\label{eq:isnr}
	\sigma = \frac{\| \boldsymbol{\Phi} \boldsymbol{x}\|_{\ell_2}}{\sqrt{M}} \cdot 10^{\frac{-\text{ISNR}}{20}}.
\end{equation}

\subsection{Training and transfer learning}\label{sec:training}
We train the networks on pairs $(\boldsymbol{y}_i, \boldsymbol{x}_i)$ of synthetic noisy \mbox{SPIDER} Fourier measurements and images with the mean squared error (MSE) cost
\begin{equation}
	\mathcal{C}(\boldsymbol{y}_i, \boldsymbol{x}_i) =  \frac{\sum_{i=1}^N \|\boldsymbol{\Phi}^\dagger_\theta \boldsymbol{y}_i - \boldsymbol{x}_i \|^2}{N},
\end{equation}
where $\boldsymbol{\Phi}^\dagger_\theta:\mathbb{C}^M \rightarrow \mathbb{R}^N$ is the learned pseudo-inverse operator (i.e. the networks defined in Section~\ref{sec:solving-approaches}), such that $\boldsymbol{\Phi}^\dagger_\theta \boldsymbol{y}$ gives the learned reconstructions of the noisy measurements $\boldsymbol{y}$.

The networks are trained using the ADAM optimizer \citep{kingmaAdamMethodStochastic2014} with a learning rate of $0.001$ and a batch size of 5. The training data for each epoch is precomputed to save time. Networks are trained for 200 epochs.

\subsection{Computation time}

The average evaluation time required to compute reconstructed images for the different methods is shown in Table~\ref{tab:times} for images of the COCO dataset. The evaluation times include the initial application of the NUFFT pseudo-inverse. Training time is also included and does not include the time of pre-computing the augmented training data.

As expected, the U-Net model is highly computationally efficient due to the small number of operator evaluations.  The GU-Net model is moderately slower than the U-Net model, but not significantly so, again as expected.  We note that learned methods are trained and evaluated on the GPU while the primal-dual method is run on a CPU, precluding direct comparison of the reconstruction time.  Nevertheless, the evaluation times in Table 3 are roughly proportional to the number of fine scale operator evaluations, which corroborates that these evaluations dominate the computational cost, and so a GPU implementation of the primal-dual algorithm would remain $\sim 600\times$ slower than our U-Net model and $\sim 55\times$ slower than our GU-Net model.  The computational times required by our learned models to recover images, in the order of 10s of milliseconds, is sufficiently low that our proposed methods can indeed open up real-time imaging for the SPIDER instrument.

\begin{table*}
	\centering
	\caption{Number of full scale measurement operator calls, reconstruction time averaged over 1000 images of the COCO dataset and training time, for the pseudo-inverse, our learned methods, and a variational regularization approach. The reconstruction times for our learned methods are sufficiently low to enable real-time imaging for the SPIDER instrument.}
	\label{tab:times}
	\begin{tabular}{lccc} 
		\hline
		Name                          & Operator evaluations & Average reconstruction time (ms) & Training time (mins)                                   \\
		\hline
		Pseudo-inverse  (1 GPU)       & $1$                  & $   5.50 $                       & -                                                      \\
		U-Net        (1 GPU)          & $1$                  & $   10.7 $                       & ${\sim} 30$                                            \\
		GU-Net      (1 GPU)           & $11^*$               & $   42.1 $                       & ${\sim} 100$                                           \\
		Primal-Dual   (300its, 1 CPU) & $600$                & $4.7 \times 10^{4}  $            & -                                                      \\
		\hline
		\multicolumn{4}{p{.7\textwidth}}{${}^*$Refers to operator evaluation at the finest scale, which dominates the computational time of the GU-Net.} \\
	\end{tabular}
\end{table*}

\subsection{Reconstruction quality}
\begin{figure*}
	\includegraphics[width=0.49\textwidth]{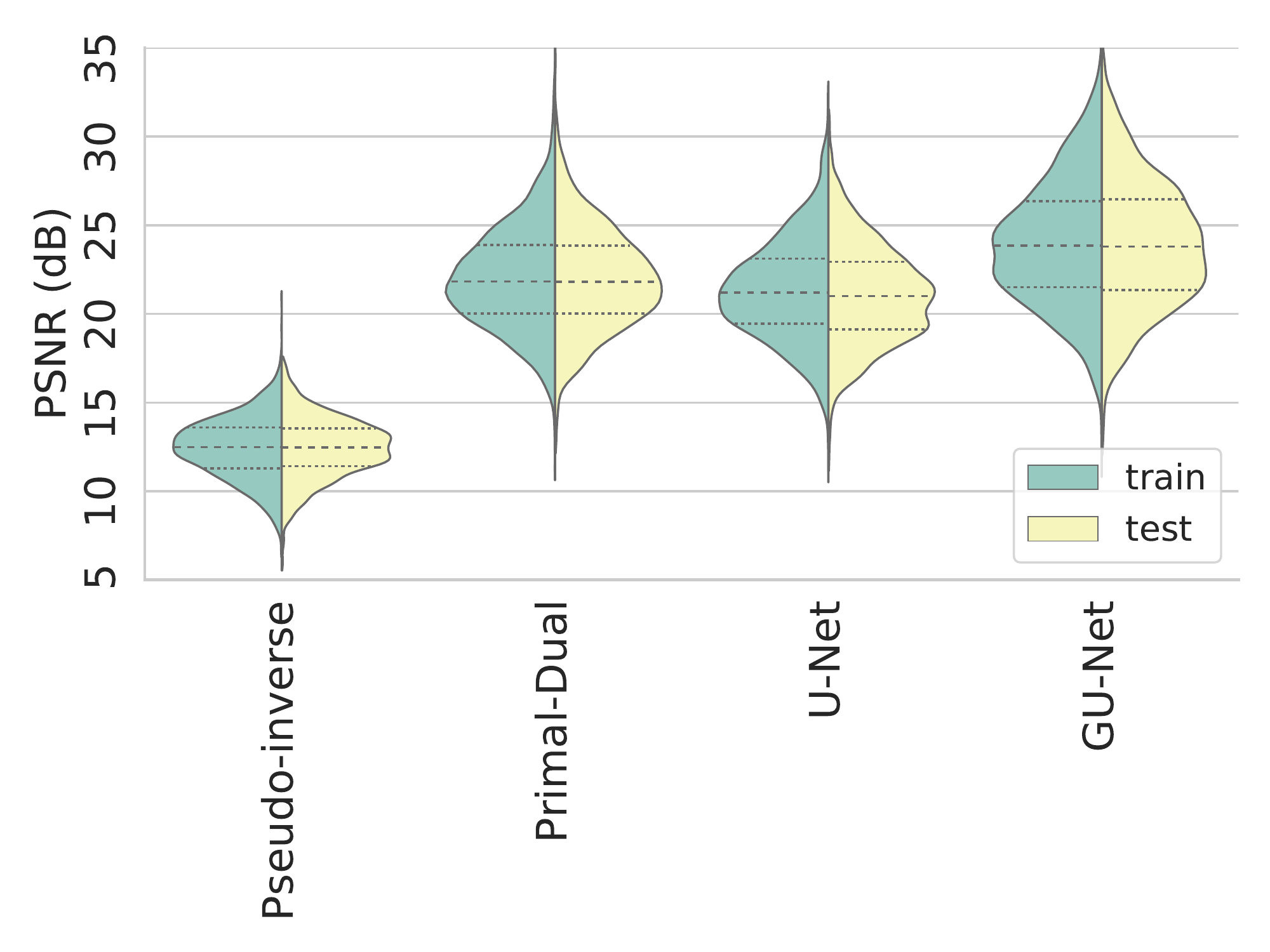}
	\includegraphics[width=0.49\textwidth]{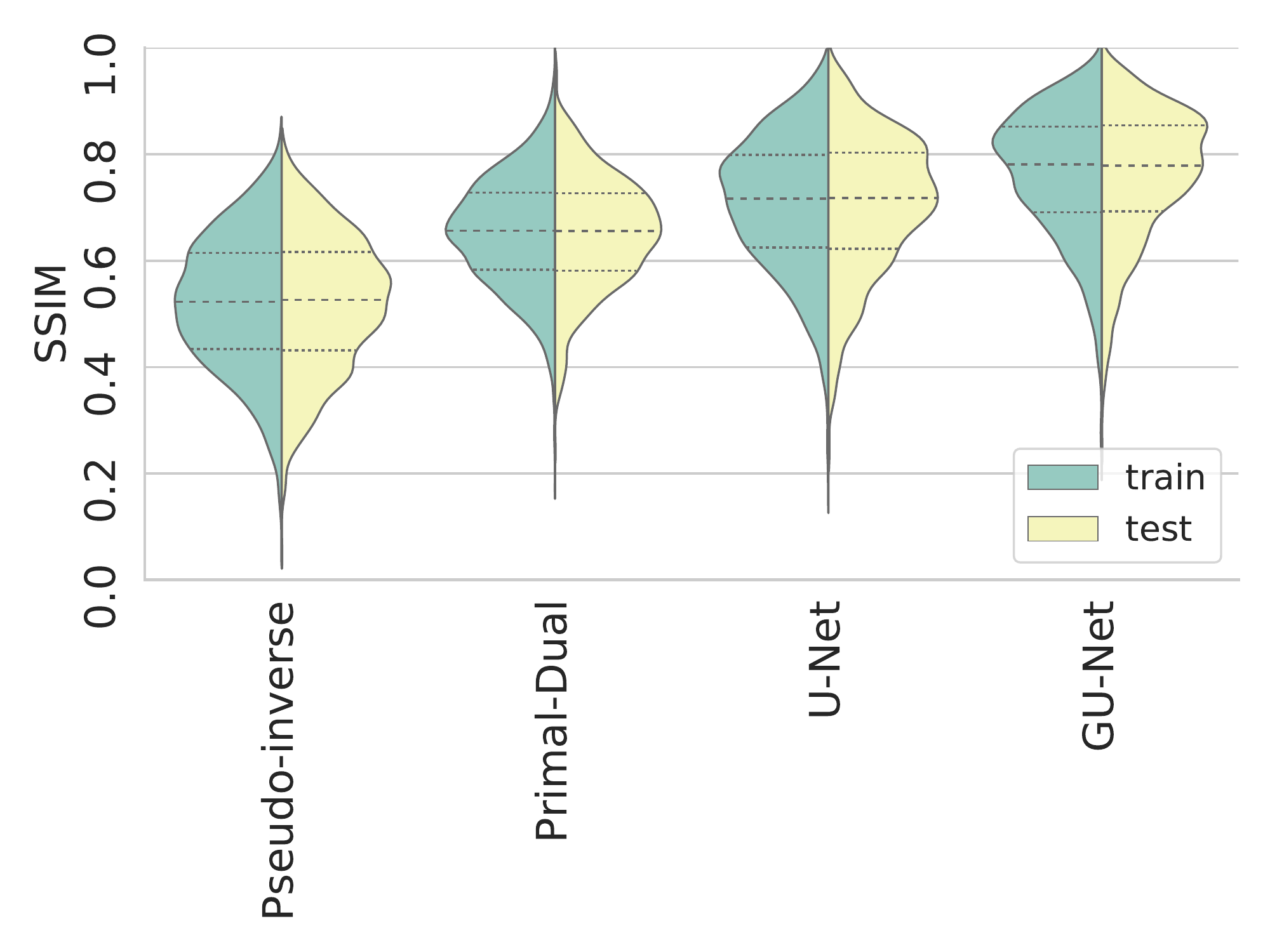}

	\caption{Distribution of reconstruction quality (PSNR and SSIM) for the different reconstruction methods on both the train and the test sets from the COCO dataset of natural images for measurements with an ISNR of 30dB. The reconstructions are made using the pseudo-inverse of the measurement operator, a primal-dual optimization scheme representing the state-of-the-art, our learned post-processing approach (U-Net), and our learned unrolled iterative approach (GU-Net). The dashed and dotted lines inside the distributions indicate the mean and quartiles of the distributions.}
	\label{fig:violin-coco}
\end{figure*}

\begin{figure*}
	\includegraphics[width=\textwidth, trim= 9.5cm 2.75cm 5.5cm 0cm, clip]{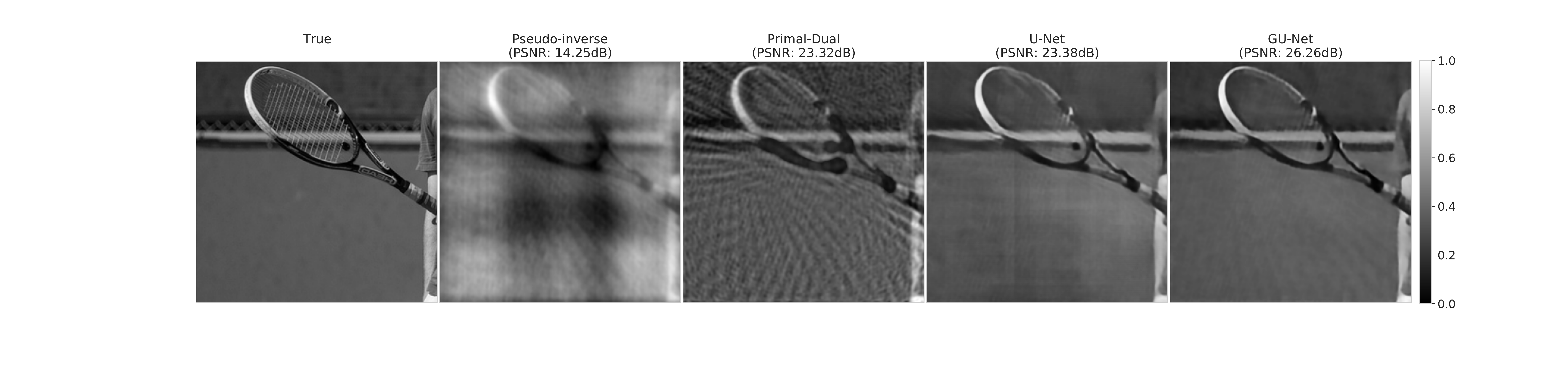}
	\includegraphics[width=\textwidth, trim= 9.5cm 2.75cm 5.5cm 2cm, clip]{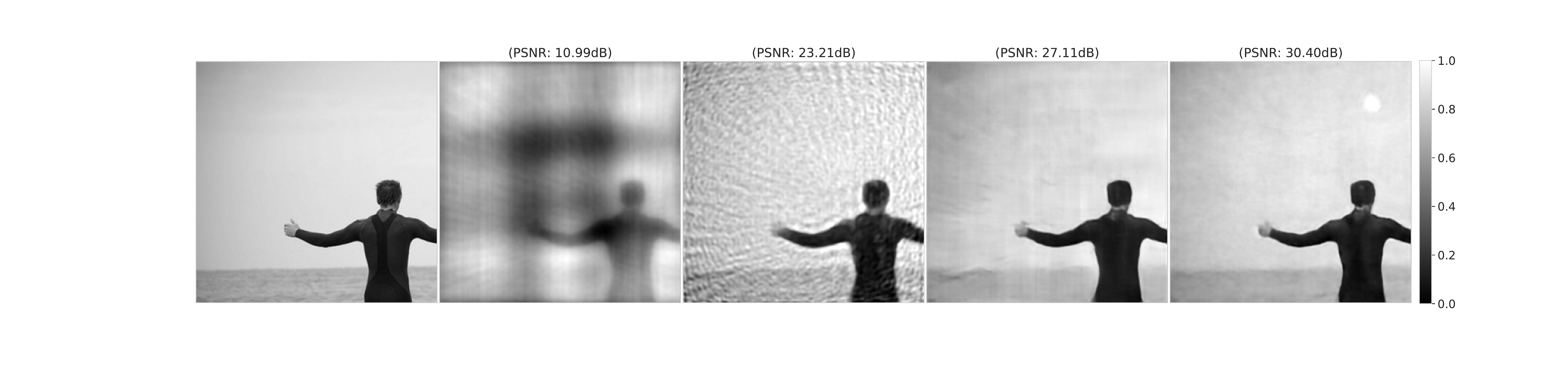}
	\includegraphics[width=\textwidth, trim= 9.5cm 2.75cm 5.5cm 2cm, clip]{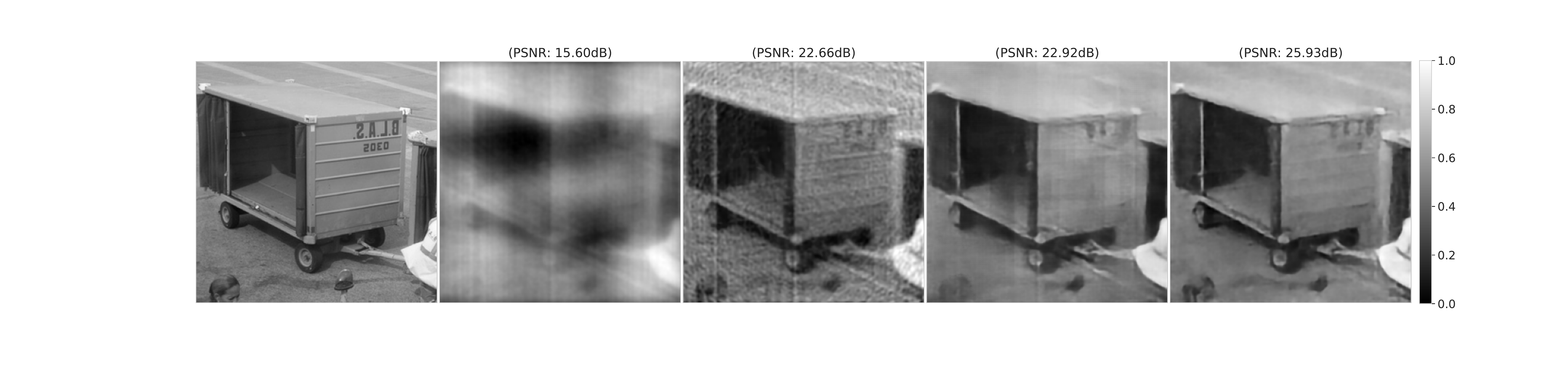}
	\includegraphics[width=\textwidth, trim= 9.5cm 2.75cm 5.5cm 2cm, clip]{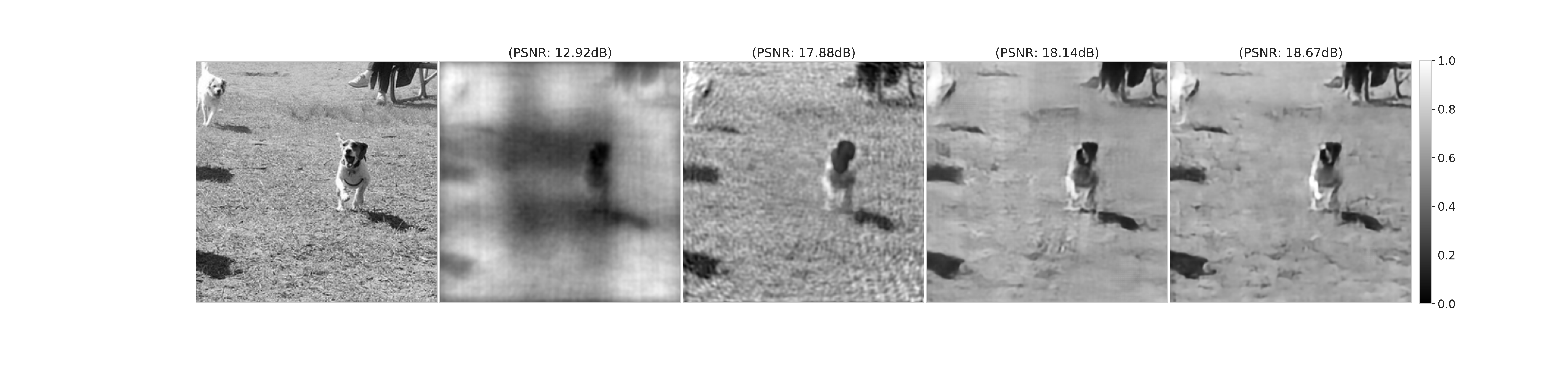}
	\includegraphics[width=\textwidth, trim= 9.5cm 2.75cm 5.5cm 2cm, clip]{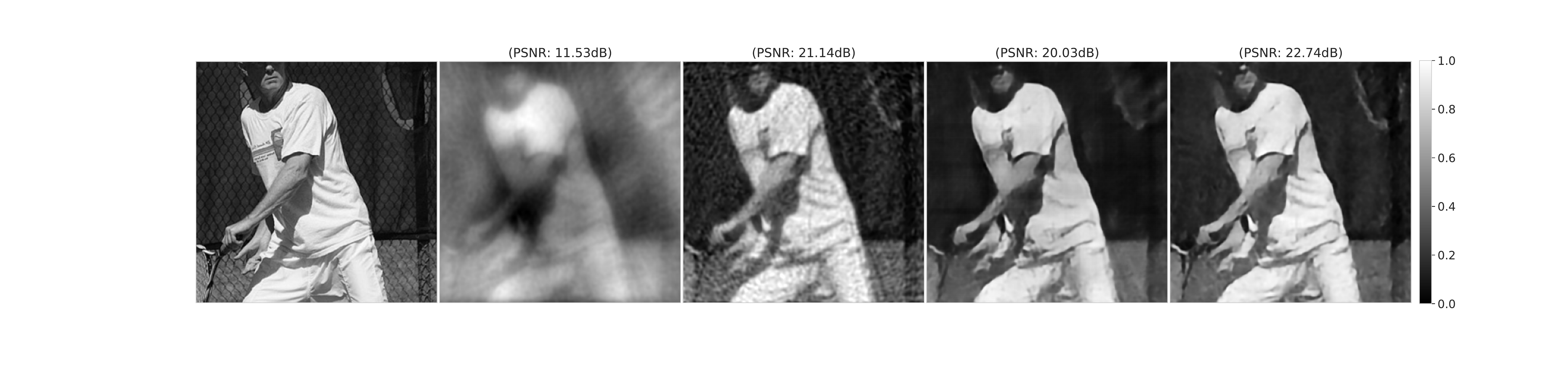}

	\caption{
		Example reconstructed images for the pseudo-inverse, the primal-dual algorithm, our learned post-processing approach (U-Net), and our learned unrolled iterative approach (GU-Net) on five images from the COCO test set. The first column shows the true image followed by the four different reconstructions from  noisy measurements with an ISNR of 30dB.}
	\label{fig:example-coco}
\end{figure*}

We first compare the pseudo-inverse, primal-dual, and two learned reconstruction methods on the COCO dataset. The distribution of the quantitative metrics (PSNR and SSIM) over the training and test sets are depicted in Figure~\ref{fig:violin-coco}. The reconstructions for a subset of training and test images from COCO dataset are illustrated in Figure~\ref{fig:example-coco}.

Our U-Net model performs similarly to the primal-dual algorithm, which represents the state-of-the-art variational regularization approach, with marginally lower PSNR but marginally greater SSIM.  However, recall that imaging with the U-Net model is orders of magnitude faster than the primal-dual algorithm.  Our GU-Net model outperforms both the U-Net and primal-dual approaches in both PSNR and SSIM, as expected since it makes greater use of knowledge of the measurement operator, while introducing only a moderate increase in computation time compared to the U-Net approach.   All methods outperform the pseudo-inverse, which is to be expected.

We note the near mirror symmetry of the plots between training and test set, which indicates only small difference in distribution of the metrics between the train and test sets (this is also the case for the primal-dual method which is dataset agnostic since it does not include any training). This suggests that the trained models generalize well to unseen data in the same domain.

From the examples in Figure~\ref{fig:example-coco} it is apparent that the pseudo-inverse yields noisy reconstructions with aliasing artefacts, the primal-dual algorithm provides an improvement but still yields reconstructions with considerable artefacts, the U-Net model generates good reconstructions that are sometimes over-smoothened, losing details, while the GU-Net model does best at recovering details and suppressing artefacts, resulting in the overall best reconstructions.

\subsection{Robustness to noise}
\begin{figure*}
	\centering
	\includegraphics[width=\textwidth]{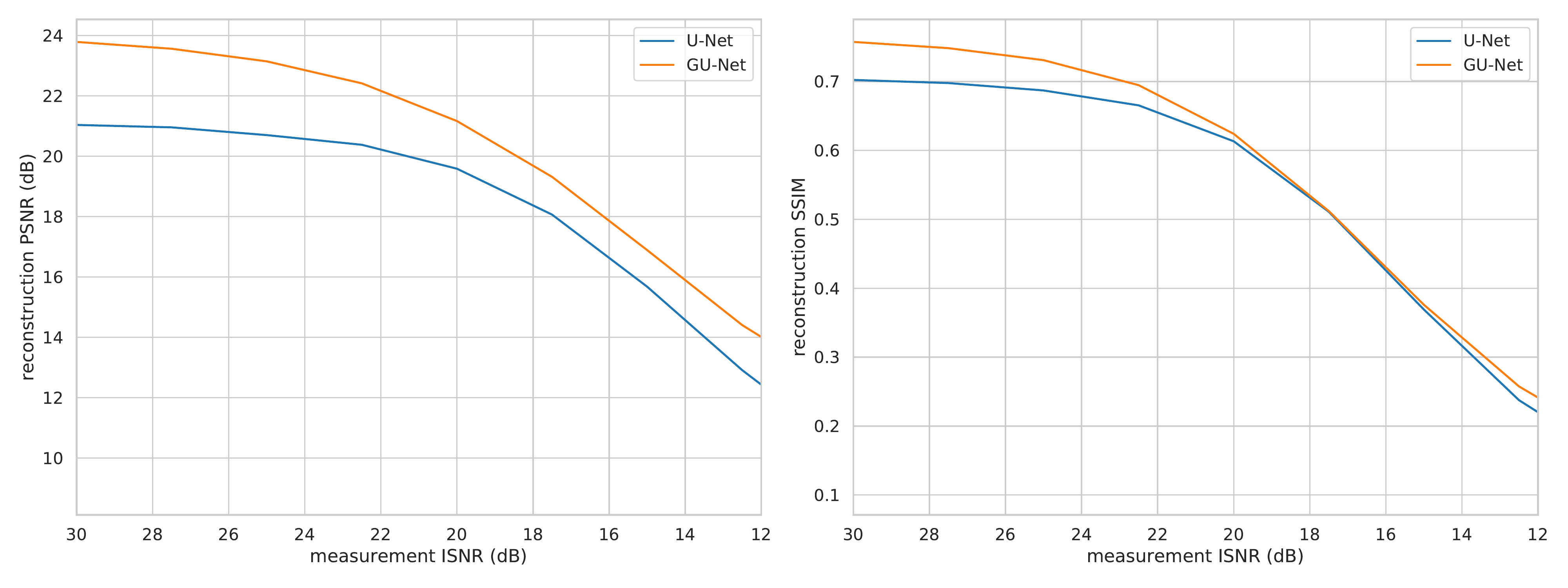}
	\caption{
		Robustness of learned models (trained for $\text{ISNR} = 30\text{dB}$) to increased input noise level: mean PSNR (left), SSIM (right) averaged over a set of 100 images from the COCO test set as a function of ISNR in the noise model (as described in Section~\ref{sec:simulating-measurements}). Our learned models exhibit sufficient robustness to variations in noise, such that a moderate underestimation of the noise level will not result in a significant loss of reconstruction quality.
	}
	\label{fig:robustness}
\end{figure*}

Next, we evaluate the robustness of our trained models with respect to increased levels of additive Gaussian noise in the input. All models are trained with $\text{ISNR} = 30\text{dB}$, however we vary the ISNR from 30dB to 12.5dB in test images. The averages of PSNR and SSIM over a set of COCO images are shown in Figure~\ref{fig:robustness}.

The plots in Figure~\ref{fig:robustness} both show a relative plateau followed by a transition to linear decay at around $\text{ISNR} = 20\text{dB}$ indicating robustness to a moderate increase in noise level, with a linear deprecation for higher noise levels. The PSNR GU-Net curve remains above the U-Net curve as noise increases, while the SSIM curves are essentially on top of one another for higher noise levels. This suggests that the advantage of the GU-Net in terms of SSIM is marginal for higher noise levels, while we maintain some PSNR advantage even for higher noise levels.
Overall, Figure~\ref{fig:robustness} suggest sufficient robustness of the trained models with respect to varying input noise levels, such that a moderate underestimation of the noise level will not result in a significant loss of reconstruction quality.

\subsection{Generalization to other datasets}

Besides assessing the performance of the models on natural images, where we have a reasonably large number of training instances, we also consider other datasets where data availability is limited, as is the case for the galaxy and satellite datasets described in Section~\ref{sec:datasets}.

We consider three approaches for training networks for the two smaller, domain-specific datasets: (i) training on the small dataset from scratch;  (ii) reconstruction of test images using the network trained on the COCO dataset without retraining; and (iii) transfer learning via initializing with the COCO pretrained networks and fine-tuning on images from the small dataset for 100 epochs.

\begin{figure*}
	\includegraphics[width=\textwidth, trim= 7.5cm 1.5cm 4cm 0cm, clip]{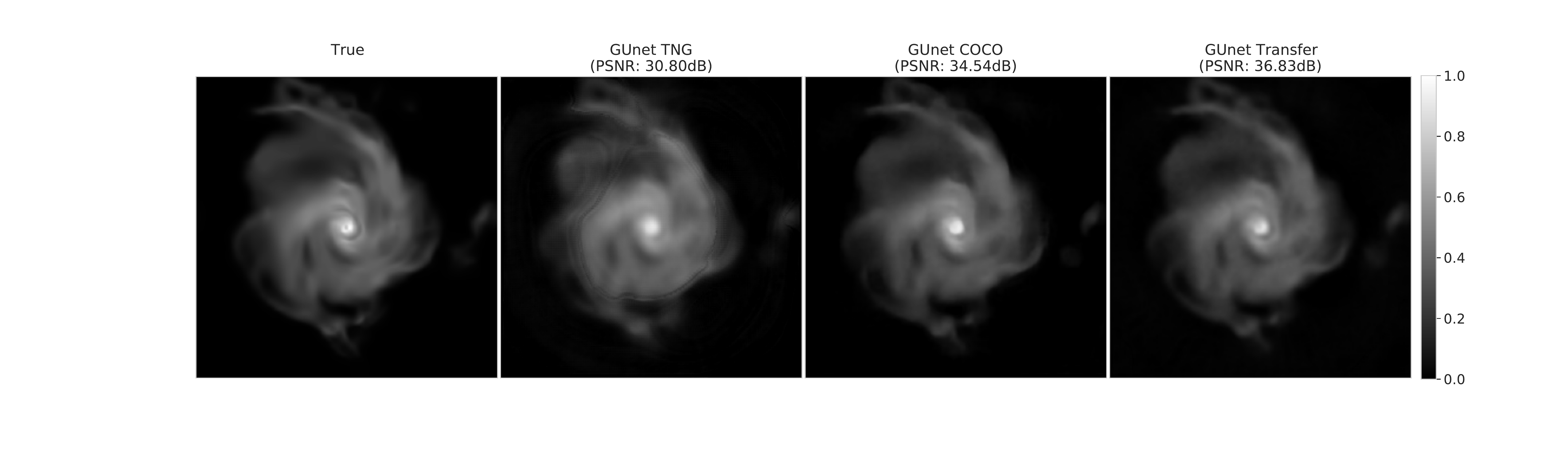}
	\caption{Comparison of the three different approaches for training a model for smaller, domain-specific datasets (as described in Section~\ref{sec:training}). From left to right: the original image, a reconstruction from a network trained only on the small galaxy dataset, a reconstruction using a network trained only on natural images from the COCO dataset, and reconstruction made with a network trained first on natural images and then fine-tuned using transfer learning to the galaxy dataset. The transfer learning approach recovers images with the highest reconstruction quality and the least amount of artefacts.
	}
	\label{fig:example-compare-transfer}
\end{figure*}

As a first test, we compare these three approaches for the smaller, domain-specific galaxy dataset. Figure \ref{fig:example-compare-transfer} shows GU-Net reconstructions for a test galaxy image in these three scenarios.
The network trained on just the galaxy images produces artefacts in its reconstruction, due to the small volume of training data, while the model trained solely on images from the COCO dataset appears blurred, as the galaxy images do not have sharp edges. The transfer learning approach, which leverages both the COCO dataset and the smaller dataset of galaxy images, recovers the image with the highest PSNR. Therefore, we restrict further comparisons to the transfer leaning scenario.

\begin{figure*}
	\includegraphics[width=.49\textwidth]{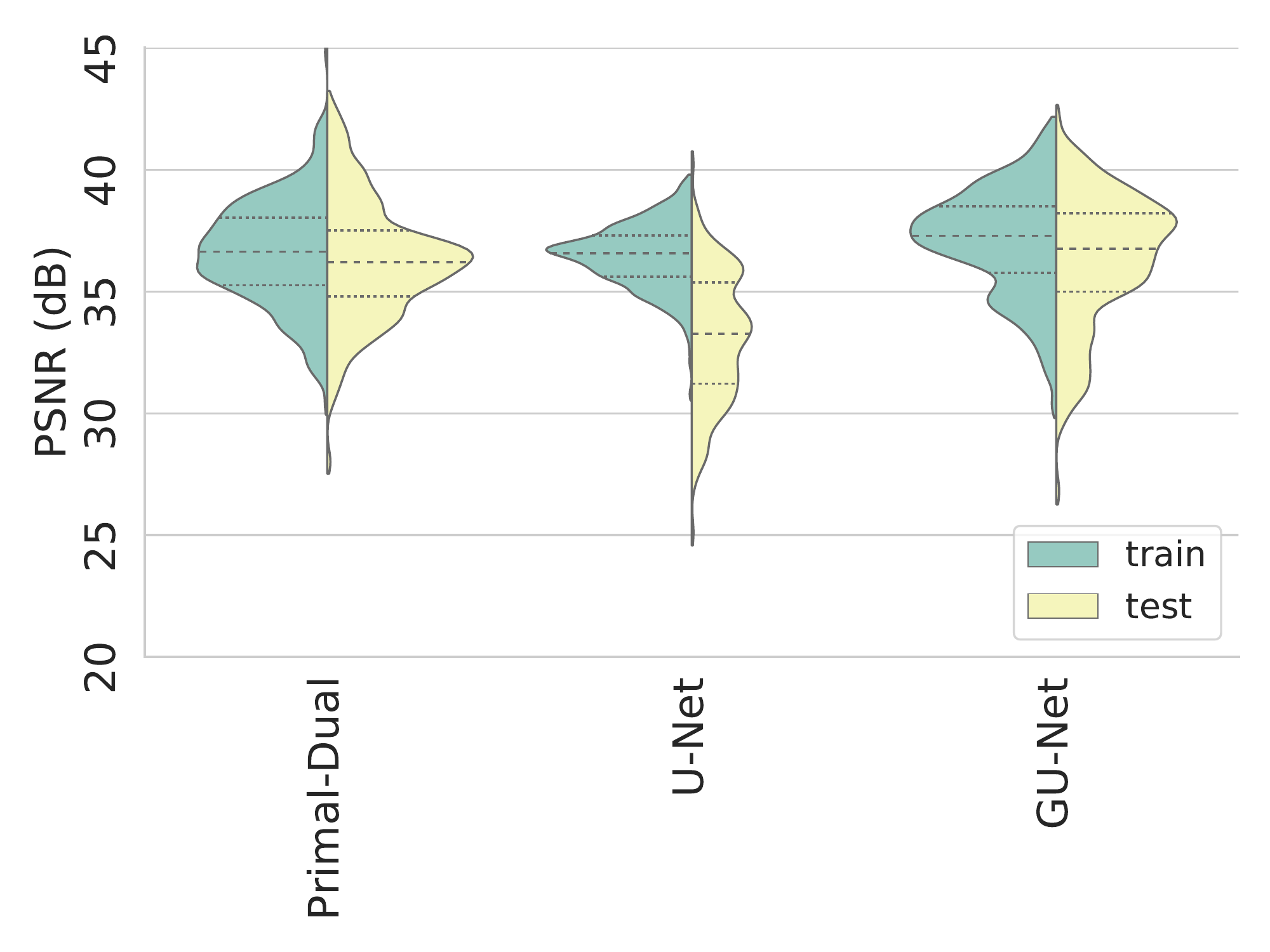}
	\includegraphics[width=.49\textwidth]{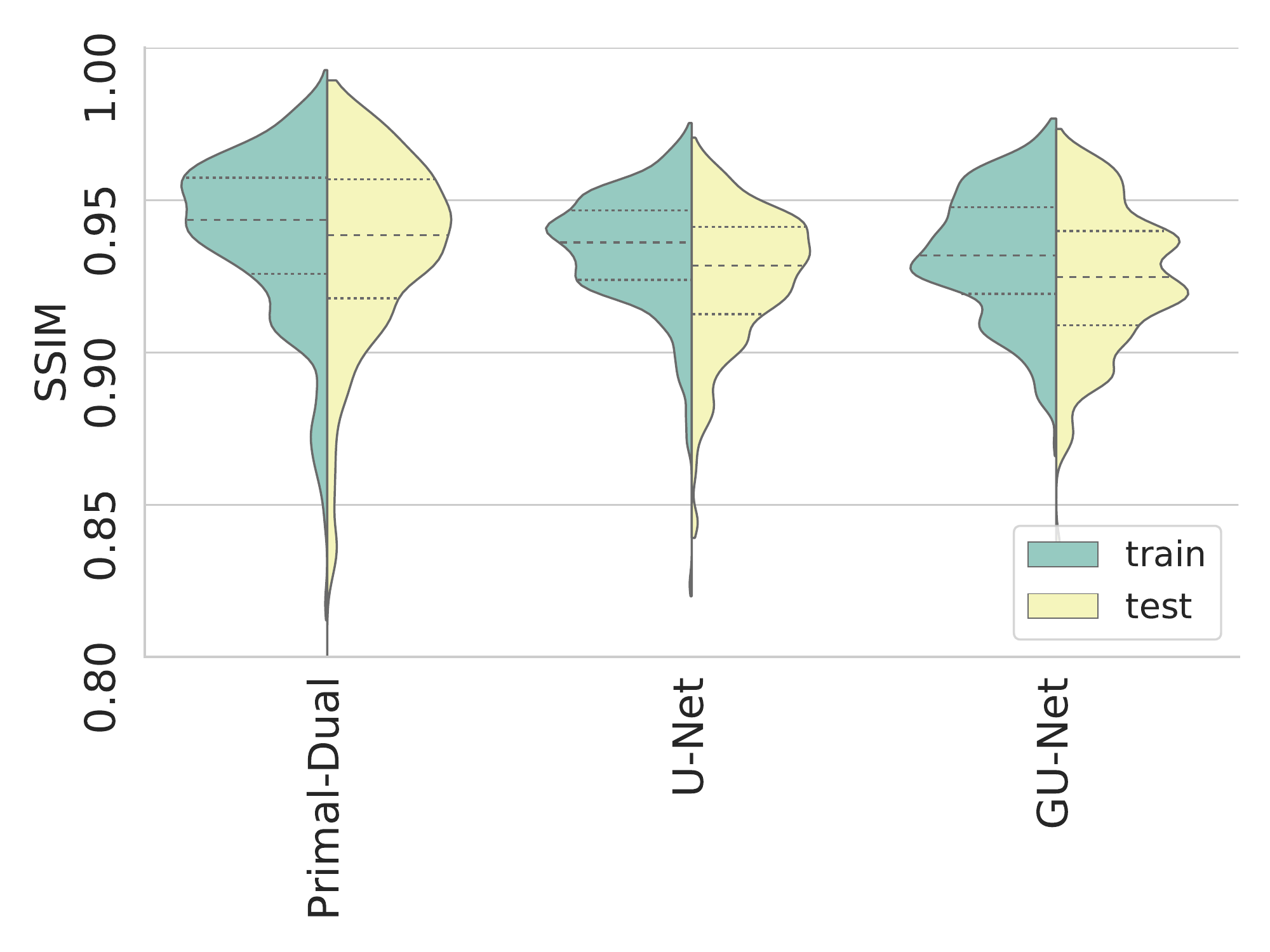}

	\caption{
		Distribution of quantitive imaging metrics (PSNR and SSIM) over the train and test sets of the galaxy dataset. The dashed and dotted lines indicate the mean and quartiles of the distributions. The learned approaches are trained using images of the COCO dataset first, and are then adapted to the galaxy dataset by transfer learning. }
	\label{fig:violin-tng}
\end{figure*}

\begin{figure*}
	\includegraphics[width=\textwidth, trim= 9.5cm 2.75cm 5.5cm 0cm, clip]{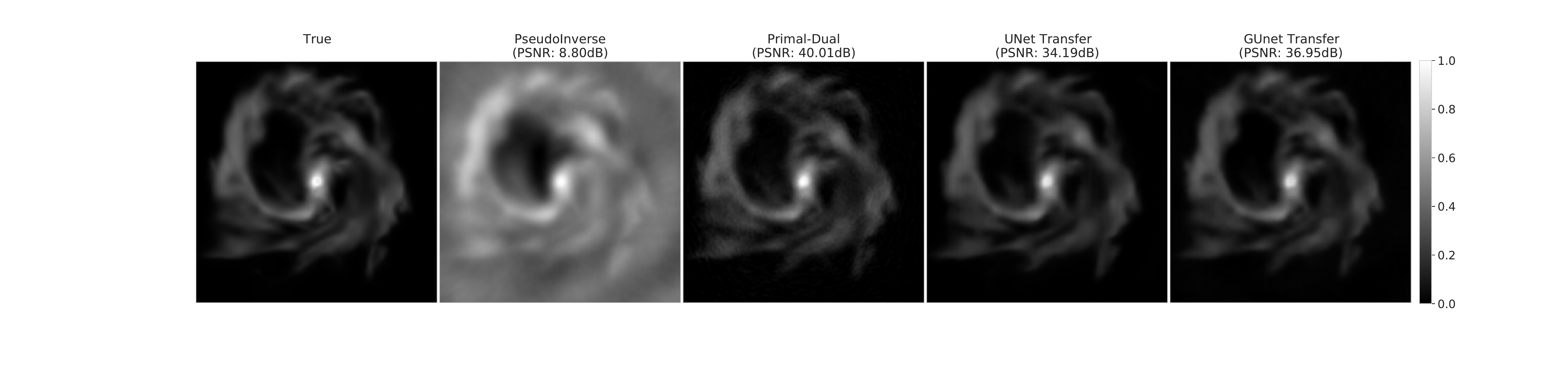}
	\includegraphics[width=\textwidth, trim= 9.5cm 2.75cm 5.5cm 2cm, clip]{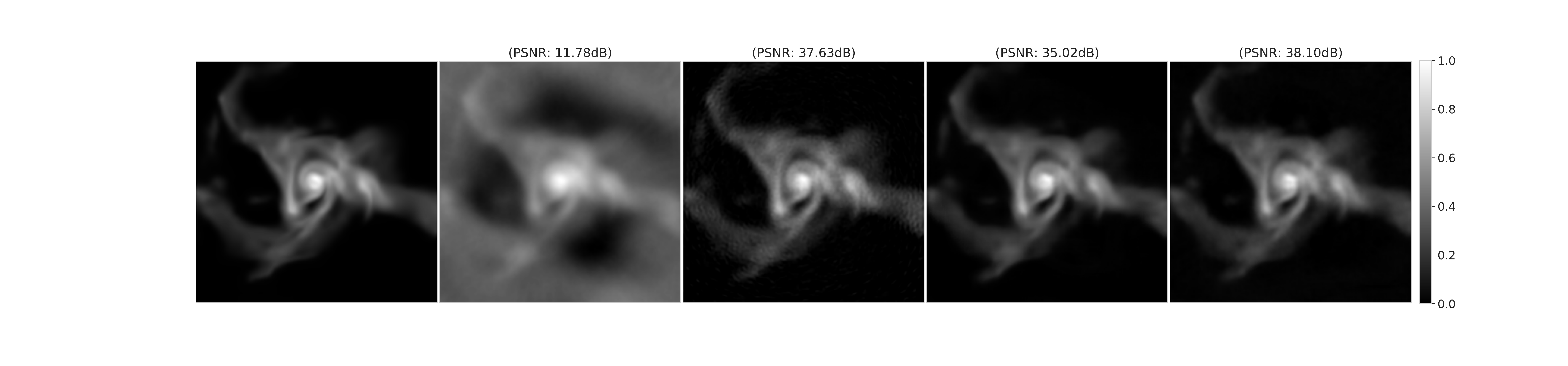}
	\includegraphics[width=\textwidth, trim= 9.5cm 2.75cm 5.5cm 2cm, clip]{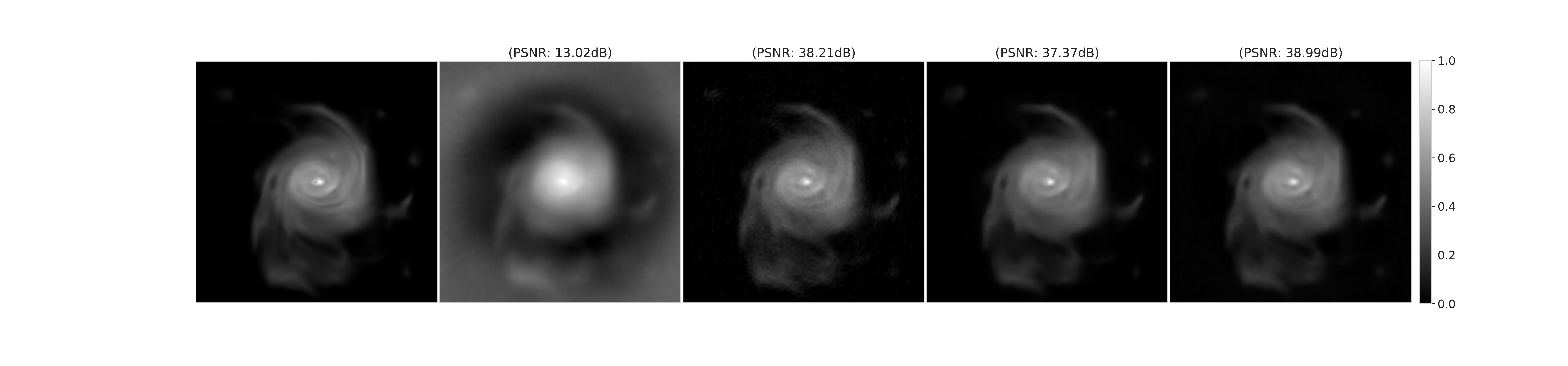}
	\includegraphics[width=\textwidth, trim= 9.5cm 2.75cm 5.5cm 2cm, clip]{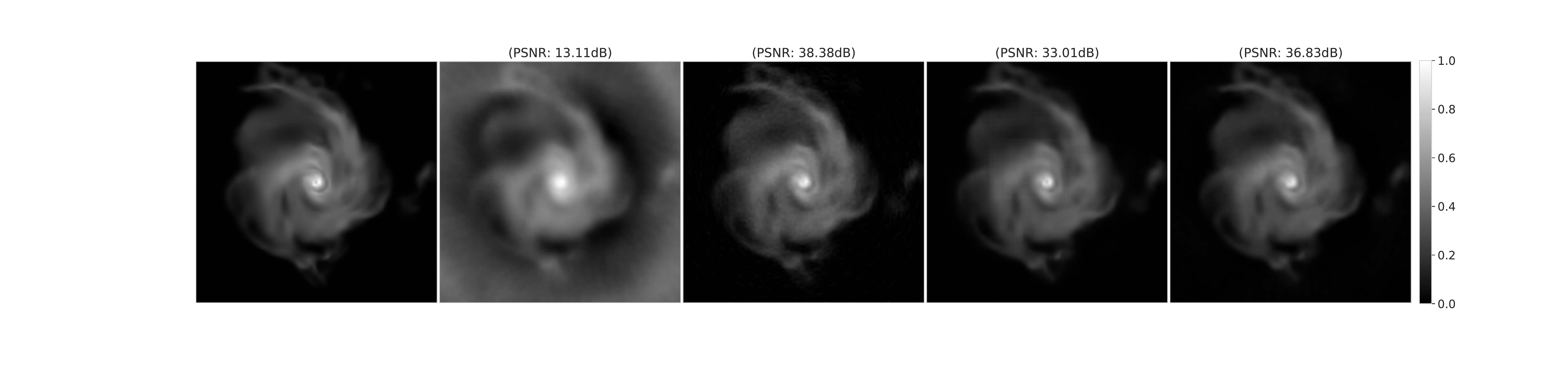}
	\includegraphics[width=\textwidth, trim= 9.5cm 2.75cm 5.5cm 2cm, clip]{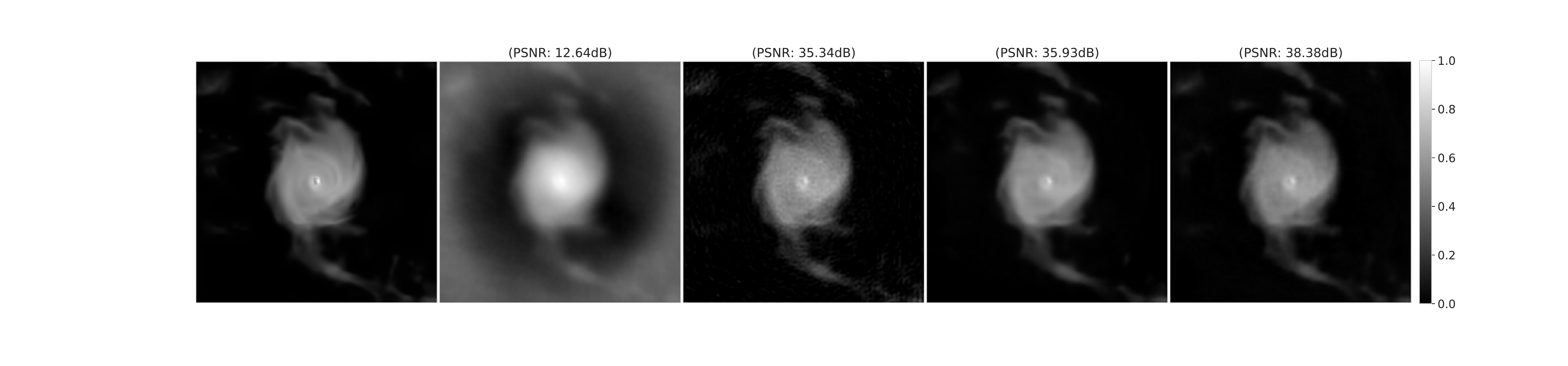}

	\caption{Reconstruction performance of the pseudo-inverse, a state-of-the-art reconstruction (primal-dual), our learned post-processing approach (U-Net), and our learned unrolled iterative approach (GU-Net) on five images from the galaxy images test set. The first column shows the true image followed by the four different reconstructions from noisy measurements with an ISNR of 30dB. The learned methods are trained on images from the COCO dataset first and are then adapted to the galaxy dataset by transfer learning. }
	\label{fig:example-TNG}
\end{figure*}

The distribution of the quantitative metrics over the training and test sets of the galaxy dataset for the transfer learning scenario can be seen in Figure~\ref{fig:violin-tng}. Example reconstructions can be seen in Figure~\ref{fig:example-TNG}.
From Figure~\ref{fig:violin-tng} it is evident that the U-Net has been over-fitted to the training data as its performance metrics on the test data are substantially worse.  Nevertheless, the example reconstructions of both learned models in Figure\ref{fig:example-TNG} are visually very similar. We suppose this is due to the nature of the galaxy dataset for which neither PSNR nor SSIM is a particularly well suited metric. Indeed, visually the galaxy images reconstructed by the learned models are arguably slightly more appealing that those recovered by the primal-dual algorithm, while the quantitative metrics are similar or arguably marginally favour the primal-dual case.  The close match between the train and test distributions for GU-Net indicates that incorporating the physical model into the learned reconstruction yields models with superior generalization performance to unseen data.

\begin{figure*}
	\includegraphics[width=.49\textwidth]{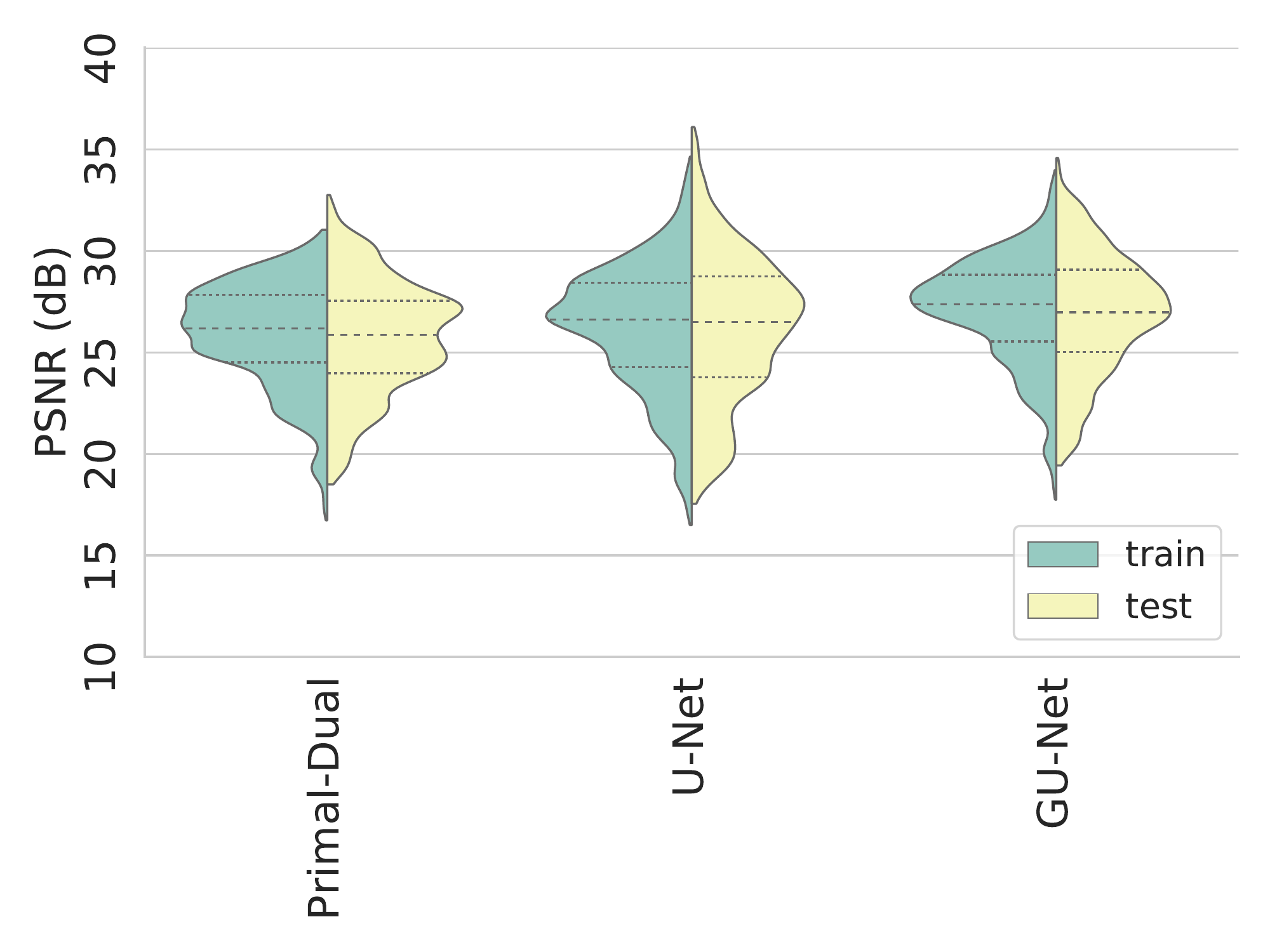}
	\includegraphics[width=.49\textwidth]{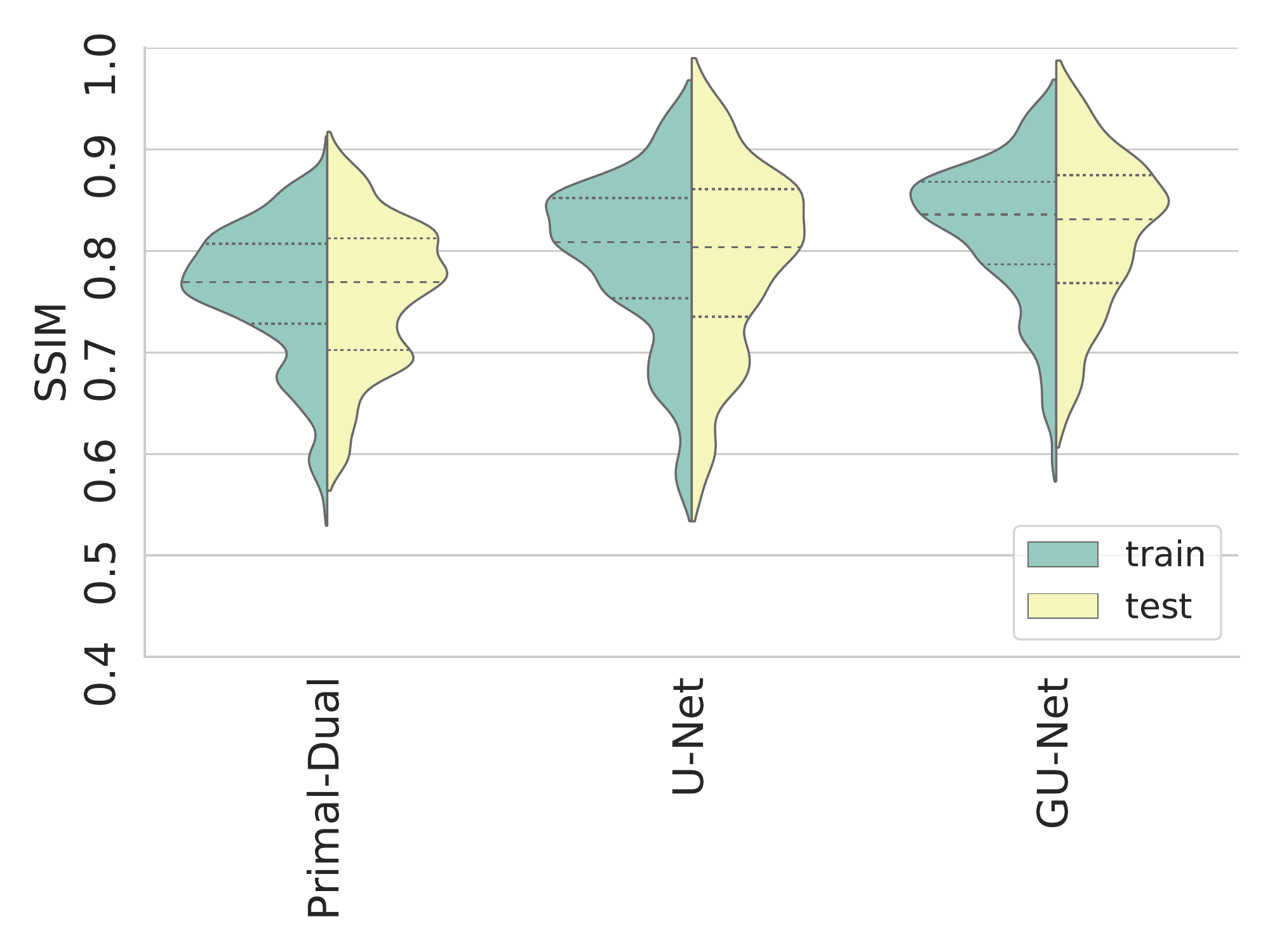}

	\caption{Distribution of quantitive imaging metrics (PSNR and SSIM) over the train and test sets of the Deep Globe satellite image dataset. The dashed and dotted lines indicate the mean and quartiles of the distributions. The learned approaches are trained using images of the COCO dataset first, and are then adapted to the satellite image dataset by transfer learning.
	}
	\label{fig:violin-sats}
\end{figure*}

\begin{figure*}
	\includegraphics[width=\textwidth, trim= 9.5cm 2.75cm 5.5cm 0cm, clip]{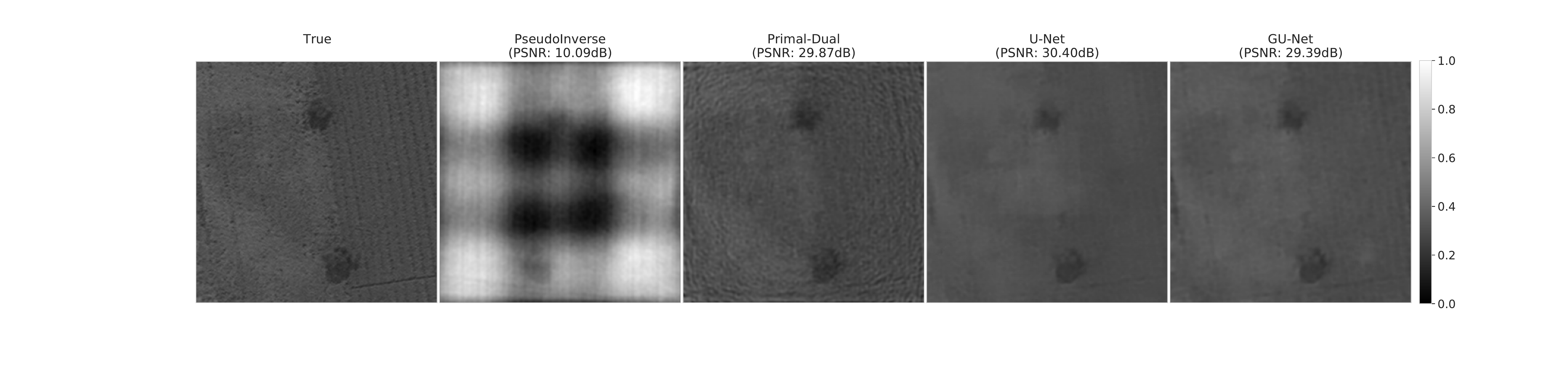}
	\includegraphics[width=\textwidth, trim= 9.5cm 2.75cm 5.5cm 2cm, clip]{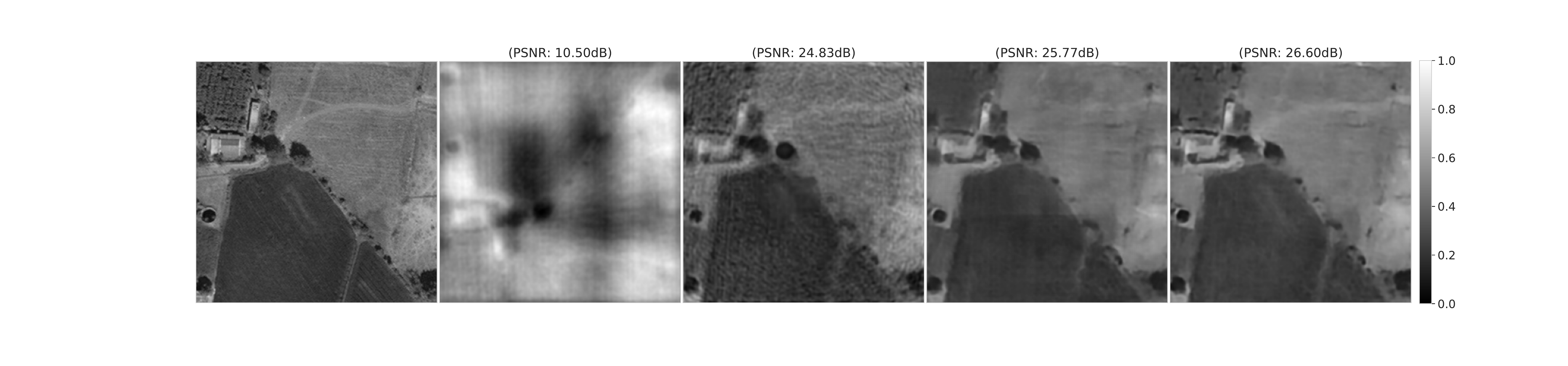}
	\includegraphics[width=\textwidth, trim= 9.5cm 2.75cm 5.5cm 2cm, clip]{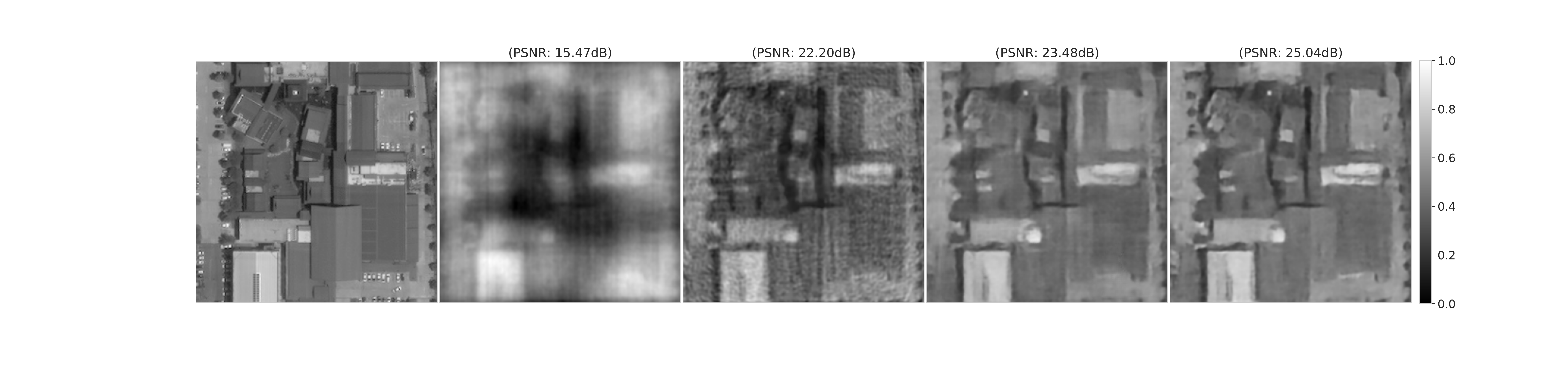}
	\includegraphics[width=\textwidth, trim= 9.5cm 2.75cm 5.5cm 2cm, clip]{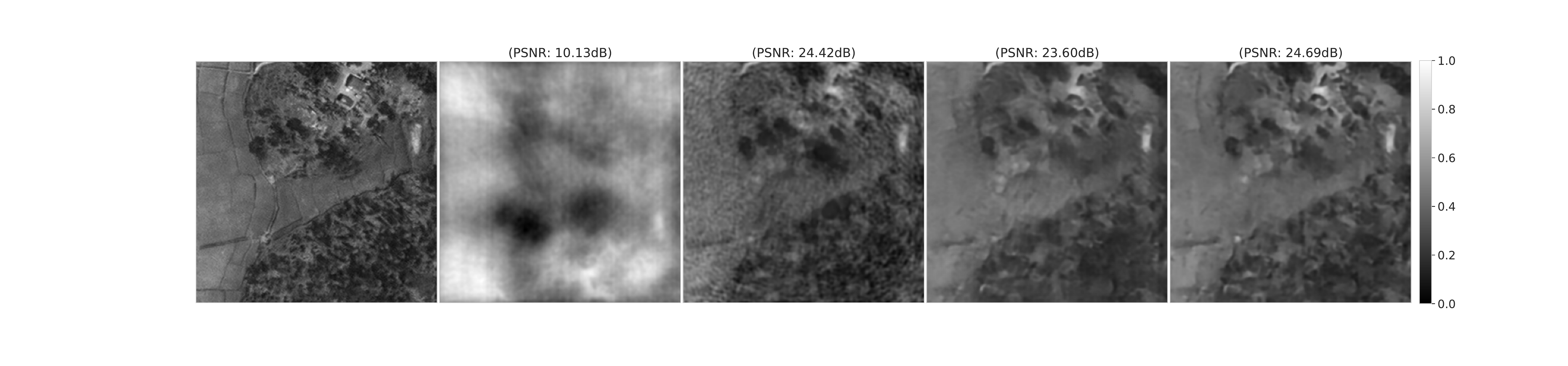}
	\includegraphics[width=\textwidth, trim= 9.5cm 2.75cm 5.5cm 2cm, clip]{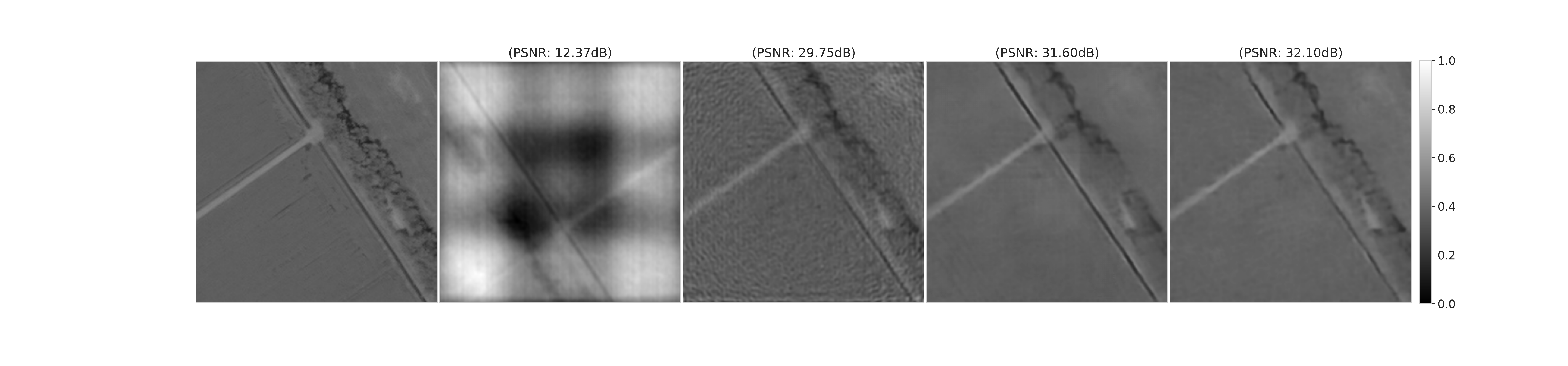}

	\caption{Reconstruction performance of the pseudo-inverse, a state-of-the-art reconstruction (primal-dual), our learned post-processing approach (U-Net), and our learned unrolled iterative approach (GU-Net) on five images from the satellite images test set. The first column shows the true image followed by the four different reconstructions from  noisy measurements with an ISNR of 30dB. The learned methods are trained on images from the COCO dataset first and are then adapted to the satellite image dataset by transfer learning.}
	\label{fig:example-sats}
\end{figure*}

The same experiments are performed for the satellite image dataset, with distributions of metrics shown in Figure~\ref{fig:violin-sats} and example reconstructions shown in Figure~\ref{fig:example-sats}.
Both our learned approaches outperform the primal-dual approach, in terms of metrics and visual inspection (the primal-dual algorithm yields more grainy, noisy reconstructions).  Moreover, the GU-Net models exhibits a slight improvement over the U-Net model, particularly for images with clear edges.

In summary, by adopting a transfer learning approach our learned imaging methods achieve similar or superior reconstruction quality to a state-of-the-art variational regularization approach even for datasets for which limited training data is available, in a fraction of the computational time.

\section{Conclusions}\label{sec:conclusion}

We have proposed two new learned approaches to reconstruct images for the SPIDER instrument, which compared to the classical state-of-the-art proximal optimization algorithms improve reconstruction quality, while dramatically reducing the computational time required to recover images.

Our first learned method adopts a learned post-processing approach to clearly separate the physical measurement model from the learned imaging via a U-Net model.  A consequence of this configuration is that the pseudo-inverse of the physical measurement model need only be applied once, resulting in orders of magnitude reduction in imaging time compared to the benchmark traditional optimization algorithm.  Reconstruction quality is similar to the benchmark traditional algorithm.

Our second learned method, the GU-Net, uses a similar architecture but enhanced with multiscale evaluations of the gradient of the data fidelity term, interweaving the measurement model into the U-Net architecture.  While this results in a moderate increase in the time required to recover images compared to the U-Net model, reconstruction quality is improved by making greater use of knowledge of the physical measurement model.

Overall, our learned methods achieve similar or superior reconstruction quality compared to traditional approaches, while realizing a dramatic reduction in the time required to recover images, to the extent that real-time imaging with the SPIDER instrument becomes possible for the first time, opening up many new use-cases.

While our learned methods are trained using data with a specific noise level, their performance on measurements with increased noise levels remains similar for a moderate increase in noise level.
Furthermore, in scenarios where only a limited volume of training data is available, as common in domain-specific problems, the performance of the models can be increased by transfer learning from a domain with sufficient training data. While for actual observations one should try to construct as representative training as possible, our learned methods show that it is possible to achieve similar or superior reconstruction quality to the traditional primal-dual approach in settings where limited train data are available, and at a fraction of the computational cost.

In addition, we have presented two approaches to modelling the measurement process of the SPIDER instrument. One method is based on the NUFFT, which is widely used in radio interferometry and is applicable to any arbitrary sampling distribution.  We also present a new modelling approach based on the Radon transform, an NU-Radon method, that applies specifically when the sampling distribution has radial spokes of measurements, as is the case for \mbox{SPIDER}, and show similarities to works in the medical imaging domain.   This latter approach is computationally more efficient when the number of measurements is large and the calculation of the measurement operator is dominated by the (de)gridding of measurements or when the number of pixels is large for a modest number of spokes. Both modelling methods and all imaging techniques have been implemented in Python and are available for use in the LeIA\footnote{\url{https://github.com/astro-informatics/LeIA}} code.

While the methods presented in this work can be used with any arbitrary sampling distribution (when using the NUFFT), the sampling distribution we have considered is fixed. For radio interferometric imaging the sampling distribution typically varies for each observation as it depends on the location of the object in the sky as well as the length of time of the observation. Looking to the future, in order to extend the learned methods presented in this paper to radio interferometry, the networks need to be trained in such as way as to support varying sampling patterns.  This is the focus of ongoing research, for which preliminary results are encouraging.  Variants of the methods presented in this paper are therefore likely to be of use not only for SPIDER imaging but for radio interferometric imaging more generally.

\section*{Acknowledgements}
We thank Luke Pratley for general discussions and for his work on the OptimusPrimal code, and Matthew Price for providing the galaxy images created from the IllustrisTNG data and for general discussions.
Matthijs Mars is supported by the UCL Centre for Doctoral Training in Data Intensive Science (STFC Training grant ST/P006736/1).
This work is also partially supported by EPSRC (grant number EP/W007673/1).
The authors acknowledge the use of the UCL Myriad High Performance Computing Facility (Myriad@UCL), and associated support services, in the completion of this work.
This work also used computing equipment funded by the Research Capital Investment Fund (RCIF) provided by UKRI, and partially funded by the UCL Cosmoparticle Initiative.


\section*{Data Availability}
The data underlying this article will be shared on reasonable request to the corresponding author.



\bibliographystyle{rasti}
\bibliography{Bibliography/lib.bib} 

\begin{thebibliography}{74}
\expandafter\ifx\csname natexlab\endcsname\relax\def\natexlab#1{#1}\fi

\bibitem[Adler \& {\"O}ktem(2017)]{adlerSolvingIllposedInverse2017}
Adler, J. \& {\"O}ktem, O., 2017.
\newblock Solving ill-posed inverse problems using iterative deep neural
  networks, {\it Inverse Problems\/}, {\bf 33}(12), 124007.

\bibitem[Adler \& {\"O}ktem(2018)]{adlerLearnedPrimaldualReconstruction2018}
Adler, J. \& {\"O}ktem, O., 2018.
\newblock Learned {{Primal-dual Reconstruction}}, {\it IEEE Transactions on
  Medical Imaging\/}, {\bf 37}(6), 1322--1332.

\bibitem[Allam~Jr(2016)]{allamjrRadioInterferometricImage2016}
Allam~Jr, T., 2016.
\newblock {\it Radio Interferometric Image Reconstruction for the {{SKA}}:
  {{A}} Deep Learning Approach\/}, {{MSc}}. {{Thesis}}, University College
  London.

\bibitem[Arridge et~al.(2019)Arridge, Maass, {\"O}ktem, \&
  Sch{\"o}nlieb]{arridgeSolvingInverseProblems2019}
Arridge, S., Maass, P., {\"O}ktem, O., \& Sch{\"o}nlieb, C.-B., 2019.
\newblock Solving inverse problems using data-driven models, {\it Acta
  Numerica\/}, {\bf 28}, 1--174.

\bibitem[Badham et~al.(2017)Badham, Kendrick, Wuchenich, Ogden, Chriqui,
  Duncan, Thurman, Yoo, Su, Lai, Chun, Li, \&
  Liu]{badhamPhotonicIntegratedCircuitbased2017}
Badham, K., Kendrick, R.~L., Wuchenich, D., Ogden, C., Chriqui, G., Duncan, A.,
  Thurman, S.~T., Yoo, S. J.~B., Su, T., Lai, W., Chun, J., Li, S., \& Liu, G.,
  2017.
\newblock Photonic integrated circuit-based imaging system for {{SPIDER}}, in
  {\em 2017 {{Conference}} on {{Lasers}} and {{Electro-Optics Pacific Rim}}
  ({{CLEO-PR}})\/}, pp. 1--5, {IEEE}, {Singapore, Singapore}.

\bibitem[Bobin et~al.(2007)Bobin, Starck, Fadili, Moudden, \&
  Donoho]{bobinMorphologicalComponentAnalysis2007}
Bobin, J., Starck, J.-L., Fadili, J., Moudden, Y., \& Donoho, D., 2007.
\newblock Morphological {{Component Analysis}}: {{An Adaptive Thresholding
  Strategy}}, {\it IEEE Transactions on Image Processing\/}, {\bf 16}(11),
  2675--2681.

\bibitem[Boink et~al.(2020)Boink, Manohar, \&
  Brune]{boinkPartiallyLearnedAlgorithmJoint2020}
Boink, Y.~E., Manohar, S., \& Brune, C., 2020.
\newblock A {{Partially-Learned Algorithm}} for {{Joint Photo-acoustic
  Reconstruction}} and {{Segmentation}}, {\it IEEE Transactions on Medical
  Imaging\/}, {\bf 39}(1), 129--139.

\bibitem[Boyd \& Vandenberghe(2004)]{boydConvexOptimization2004}
Boyd, S.~P. \& Vandenberghe, L., 2004.
\newblock {\it Convex Optimization\/}, {Cambridge University Press},
  {Cambridge, UK ; New York}.

\bibitem[Cai et~al.(2018{\natexlab{a}})Cai, Pereyra, \&
  McEwen]{caiUncertaintyQuantificationRadio2018}
Cai, X., Pereyra, M., \& McEwen, J.~D., 2018{\natexlab{a}}.
\newblock Uncertainty quantification for radio interferometric imaging: {{II}}.
  {{MAP}} estimation, {\it Monthly Notices of the Royal Astronomical
  Society\/}, {\bf 480}(3), 4170--4182.

\bibitem[Cai et~al.(2018{\natexlab{b}})Cai, Pereyra, \&
  McEwen]{caiUncertaintyQuantificationRadio2018a}
Cai, X., Pereyra, M., \& McEwen, J.~D., 2018{\natexlab{b}}.
\newblock Uncertainty quantification for radio interferometric imaging: {{I}}.
  proximal {{MCMC}} methods, {\it Monthly Notices of the Royal Astronomical
  Society\/}, {\bf 480}(3), 4154--4169.

\bibitem[Cai et~al.(2019)Cai, Pratley, \&
  McEwen]{caiOnlineRadioInterferometric2019}
Cai, X., Pratley, L., \& McEwen, J.~D., 2019.
\newblock Online radio interferometric imaging: Assimilating and discarding
  visibilities on arrival, {\it Monthly Notices of the Royal Astronomical
  Society\/}, {\bf 485}(4), 4559--4572.

\bibitem[Cai et~al.(2021)Cai, McEwen, \&
  Pereyra]{caiProximalNestedSampling2021}
Cai, X., McEwen, J.~D., \& Pereyra, M., 2021.
\newblock Proximal nested sampling for high-dimensional {{Bayesian}} model
  selection, {\it ArXiv e-print\/}.

\bibitem[Cand{\`e}s et~al.(2006)Cand{\`e}s, Demanet, Donoho, \&
  Ying]{candesFastDiscreteCurvelet2006}
Cand{\`e}s, E., Demanet, L., Donoho, D., \& Ying, L., 2006.
\newblock Fast {{Discrete Curvelet Transforms}}, {\it Multiscale Modeling \&
  Simulation\/}, {\bf 5}(3), 861--899.

\bibitem[Carrillo et~al.(2012)Carrillo, McEwen, \&
  Wiaux]{carrilloSparsityAveragingReweighted2012}
Carrillo, R.~E., McEwen, J.~D., \& Wiaux, Y., 2012.
\newblock Sparsity {{Averaging Reweighted Analysis}} ({{SARA}}): A novel
  algorithm for radio-interferometric imaging, {\it ArXiv e-print\/}.

\bibitem[Carrillo et~al.(2013{\natexlab{a}})Carrillo, McEwen, Van De~Ville,
  Thiran, \& Wiaux]{carrilloSparsityAveragingCompressive2013}
Carrillo, R.~E., McEwen, J.~D., Van De~Ville, D., Thiran, J.-P., \& Wiaux, Y.,
  2013{\natexlab{a}}.
\newblock Sparsity {{Averaging}} for {{Compressive Imaging}}, {\it IEEE Signal
  Processing Letters\/}, {\bf 20}(6), 591--594.

\bibitem[Carrillo et~al.(2013{\natexlab{b}})Carrillo, McEwen, \&
  Wiaux]{carrilloPURIFYNewApproach2013}
Carrillo, R.~E., McEwen, J.~D., \& Wiaux, Y., 2013{\natexlab{b}}.
\newblock {{PURIFY}}: A new approach to radio-interferometric imaging, {\it
  ArXiv e-print\/}.

\bibitem[Chambolle \& Pock(2011)]{chambolleFirstOrderPrimalDualAlgorithm2011}
Chambolle, A. \& Pock, T., 2011.
\newblock A {{First-Order Primal-Dual Algorithm}} for {{Convex Problems}} with
  {{Applications}} to {{Imaging}}, {\it Journal of Mathematical Imaging and
  Vision\/}, {\bf 40}(1), 120--145.

\bibitem[Chen et~al.(2017)Chen, Zhang, Kalra, Lin, Chen, Liao, Zhou, \&
  Wang]{chenLowDoseCTResidual2017}
Chen, H., Zhang, Y., Kalra, M.~K., Lin, F., Chen, Y., Liao, P., Zhou, J., \&
  Wang, G., 2017.
\newblock Low-{{Dose CT With}} a {{Residual Encoder-Decoder Convolutional
  Neural Network}}, {\it IEEE Transactions on Medical Imaging\/}, {\bf 36}(12),
  2524--2535.

\bibitem[Chu et~al.(2017)Chu, Shen, Yuan, \&
  Gong]{chuNumericalSimulationOptimal2017}
Chu, Q., Shen, Y., Yuan, M., \& Gong, M., 2017.
\newblock Numerical simulation and optimal design of {{Segmented Planar Imaging
  Detector}} for {{Electro-Optical Reconnaissance}}, {\it Optics
  Communications\/}, {\bf 405}, 288--296.

\bibitem[Combettes \& Pesquet(2011)]{combettesProximalSplittingMethods2011}
Combettes, P.~L. \& Pesquet, J.-C., 2011.
\newblock Proximal {{Splitting Methods}} in {{Signal Processing}}, in {\em
  Fixed-{{Point Algorithms}} for {{Inverse Problems}} in {{Science}} and
  {{Engineering}}\/}, vol.~49, pp. 185--212, eds Bauschke, H.~H., Burachik,
  R.~S., Combettes, P.~L., Elser, V., Luke, D.~R., \& Wolkowicz, H., {Springer
  New York}, {New York, NY}.

\bibitem[Connor et~al.(2022)Connor, Bouman, Ravi, \&
  Hallinan]{connorDeepRadiointerferometricImaging2022}
Connor, L., Bouman, K.~L., Ravi, V., \& Hallinan, G., 2022.
\newblock Deep radio-interferometric imaging with {{POLISH}}: {{DSA-2000}} and
  weak lensing, {\it Monthly Notices of the Royal Astronomical Society\/}, {\bf
  514}(2), 2614--2626.

\bibitem[Daubechies(1992)]{daubechiesTenLecturesWavelets1992}
Daubechies, I., 1992.
\newblock {\it Ten {{Lectures}} on {{Wavelets}}\/}, {{CBMS-NSF Regional
  Conference Series}} in {{Applied Mathematics}}, {Society for Industrial and
  Applied Mathematics}.

\bibitem[Demir et~al.(2018)Demir, Koperski, Lindenbaum, Pang, Huang, Basu,
  Hughes, Tuia, \& Raskar]{demirDeepGlobe2018Challenge2018}
Demir, I., Koperski, K., Lindenbaum, D., Pang, G., Huang, J., Basu, S., Hughes,
  F., Tuia, D., \& Raskar, R., 2018.
\newblock {{DeepGlobe}} 2018: {{A Challenge}} to {{Parse}} the {{Earth}}
  through {{Satellite Images}}, in {\em 2018 {{IEEE}}/{{CVF Conference}} on
  {{Computer Vision}} and {{Pattern Recognition Workshops}} ({{CVPRW}})\/}, pp.
  172--17209, {IEEE}, {Salt Lake City, UT}.

\bibitem[Dokmani{\'c} et~al.(2016)Dokmani{\'c}, Bruna, Mallat, \& {de
  Hoop}]{dokmanicInverseProblemsInvariant2016}
Dokmani{\'c}, I., Bruna, J., Mallat, S., \& {de Hoop}, M., 2016.
\newblock Inverse {{Problems}} with {{Invariant Multiscale Statistics}}, {\it
  ArXiv e-print\/}.

\bibitem[Dong et~al.(2016)Dong, Loy, He, \&
  Tang]{dongImageSuperResolutionUsing2016}
Dong, C., Loy, C.~C., He, K., \& Tang, X., 2016.
\newblock Image {{Super-Resolution Using Deep Convolutional Networks}}, {\it
  IEEE Transactions on Pattern Analysis and Machine Intelligence\/}, {\bf
  38}(2), 295--307.

\bibitem[Duijndam \& Schonewille(1997)]{duijndamNonuniformFastFourier1997}
Duijndam, A. \& Schonewille, M., 1997.
\newblock Nonuniform fast {{Fourier}} transform, in {\em {{SEG Technical
  Program Expanded Abstracts}} 1997\/}, pp. 1135--1138, {Society of Exploration
  Geophysicists}.

\bibitem[Duncan et~al.(2015)Duncan, Kendrick, Thurman, Wuchenich, Scott, Yoo,
  Su, Yu, Ogden, \& Proiett]{duncanSPIDERNextGeneration2015}
Duncan, A., Kendrick, R., Thurman, S., Wuchenich, D., Scott, R.~P., Yoo,
  {\relax SJB}., Su, T., Yu, R., Ogden, C., \& Proiett, R., 2015.
\newblock {{SPIDER}}: {{Next}} generation chip scale imaging sensor, in {\em
  Advanced Maui Optical and Space Surveillance Technologies Conference\/},
  p.~27.

\bibitem[Fessler \& Sutton(2003)]{fesslerNonuniformFastFourier2003}
Fessler, J. \& Sutton, B., 2003.
\newblock Nonuniform fast fourier transforms using min-max interpolation, {\it
  IEEE Transactions on Signal Processing\/}, {\bf 51}(2), 560--574.

\bibitem[Gheller \& Vazza(2021)]{ghellerConvolutionalDeepDenoising2021}
Gheller, C. \& Vazza, F., 2021.
\newblock Convolutional {{Deep Denoising Autoencoders}} for {{Radio
  Astronomical Images}}, {\it Monthly Notices of the Royal Astronomical
  Society\/}, {\bf 509}(1), 990--1009.

\bibitem[Gregor \& LeCun(2010)]{gregorLearningFastApproximations2010}
Gregor, K. \& LeCun, Y., 2010.
\newblock Learning fast approximations of sparse coding, in {\em Proceedings of
  the 27th International Conference on International Conference on Machine
  Learning\/}, {{ICML}}'10, pp. 399--406, {Omnipress}, {Madison, WI, USA}.

\bibitem[Gribonval \& Nielsen(2003)]{gribonvalSparseRepresentationsUnions2003}
Gribonval, R. \& Nielsen, M., 2003.
\newblock Sparse representations in unions of bases, {\it IEEE Transactions on
  Information Theory\/}, {\bf 49}(12), 3320--3325.

\bibitem[Hauptmann et~al.(2020)Hauptmann, Adler, Arridge, \&
  Oktem]{hauptmannMultiScaleLearnedIterative2020}
Hauptmann, A., Adler, J., Arridge, S., \& Oktem, O., 2020.
\newblock Multi-{{Scale Learned Iterative Reconstruction}}, {\it IEEE
  Transactions on Computational Imaging\/}, {\bf 6}, 843--856.

\bibitem[Hu et~al.(2021)Hu, Liu, Zhang, Feng, \&
  Liu]{huOptimalDesignSegmented2021}
Hu, H., Liu, C., Zhang, Y., Feng, Q., \& Liu, S., 2021.
\newblock Optimal design of segmented planar imaging for dense azimuthal
  sampling lens array, {\it Optics Express\/}, {\bf 29}(15), 24300.

\bibitem[Jackson et~al.(1991)Jackson, Meyer, Nishimura, \&
  Macovski]{jacksonSelectionConvolutionFunction1991}
Jackson, J., Meyer, C., Nishimura, D., \& Macovski, A., 1991.
\newblock Selection of a convolution function for {{Fourier}} inversion using
  gridding (computerised tomography application), {\it IEEE Transactions on
  Medical Imaging\/}, {\bf 10}(3), 473--478.

\bibitem[Jin et~al.(2017)Jin, McCann, Froustey, \&
  Unser]{jinDeepConvolutionalNeural2017}
Jin, K.~H., McCann, M.~T., Froustey, E., \& Unser, M., 2017.
\newblock Deep {{Convolutional Neural Network}} for {{Inverse Problems}} in
  {{Imaging}}, {\it IEEE Transactions on Image Processing\/}, {\bf 26}(9),
  4509--4522.

\bibitem[Kendrick et~al.(2013)Kendrick, Duncan, Ogden, Wilm, Stubbs, Thurman,
  Su, Scott, \& Yoo]{kendrickFlatPanelSpaceBasedSpace2013}
Kendrick, R.~L., Duncan, A., Ogden, C., Wilm, J., Stubbs, D.~M., Thurman,
  S.~T., Su, T., Scott, R.~P., \& Yoo, {\relax SJB}., 2013.
\newblock Flat-panel space-based space surveillance sensor, in {\em Advanced
  Maui Optical and Space Surveillance Technologies Conference\/}, p. E45.

\bibitem[Kingma \& Ba(2014)]{kingmaAdamMethodStochastic2014}
Kingma, D.~P. \& Ba, J., 2014.
\newblock Adam: {{A Method}} for {{Stochastic Optimization}}, {\it ArXiv
  e-print\/}.

\bibitem[Kobler et~al.(2020)Kobler, Effland, Kunisch, \&
  Pock]{koblerTotalDeepVariation2020a}
Kobler, E., Effland, A., Kunisch, K., \& Pock, T., 2020.
\newblock Total {{Deep Variation}} for {{Linear Inverse Problems}}, in {\em
  2020 {{IEEE}}/{{CVF Conference}} on {{Computer Vision}} and {{Pattern
  Recognition}} ({{CVPR}})\/}, pp. 7546--7555, {IEEE}, {Seattle, WA, USA}.

\bibitem[Li et~al.(2020)Li, Schwab, Antholzer, \&
  Haltmeier]{liNETTSolvingInverse2020}
Li, H., Schwab, J., Antholzer, S., \& Haltmeier, M., 2020.
\newblock {{NETT}}: Solving inverse problems with deep neural networks, {\it
  Inverse Problems\/}, {\bf 36}(6), 065005.

\bibitem[Lin et~al.(2014)Lin, Maire, Belongie, Bourdev, Girshick, Hays, Perona,
  Ramanan, Zitnick, \& Doll{\'a}r]{linMicrosoftCOCOCommon2014}
Lin, T.-Y., Maire, M., Belongie, S., Bourdev, L., Girshick, R., Hays, J.,
  Perona, P., Ramanan, D., Zitnick, C.~L., \& Doll{\'a}r, P., 2014.
\newblock Microsoft {{COCO}}: {{Common Objects}} in {{Context}}, {\it ArXiv
  e-print\/}.

\bibitem[Liu et~al.(2018)Liu, Wen, \& Song]{liuSystemDesignOptical2018}
Liu, G., Wen, D.-S., \& Song, Z.-X., 2018.
\newblock System design of an optical interferometer based on compressive
  sensing, {\it Monthly Notices of the Royal Astronomical Society\/}, {\bf
  478}(2), 2065--2073.

\bibitem[Liu et~al.(2019)Liu, Wen, Song, Zhang, Li, Wei, \&
  Jiang]{liuImageReconstructionEmerging2019}
Liu, G., Wen, D., Song, Z., Zhang, W., Li, Z., Wei, X., \& Jiang, T., 2019.
\newblock Image {{Reconstruction}} of an {{Emerging Optical Imager}}, in {\em
  Image and {{Graphics}}\/}, vol. 11902, pp. 359--372, eds Zhao, Y., Barnes,
  N., Chen, B., Westermann, R., Kong, X., \& Lin, C., {Springer International
  Publishing}, {Cham}.

\bibitem[Lunz et~al.(2018)Lunz, {\"O}ktem, \&
  Sch{\"o}nlieb]{lunzAdversarialRegularizersInverse2018}
Lunz, S., {\"O}ktem, O., \& Sch{\"o}nlieb, C.-B., 2018.
\newblock Adversarial {{Regularizers}} in {{Inverse Problems}}, {\it ArXiv
  e-print\/}.

\bibitem[Lv et~al.(2020)Lv, Chen, Feng, Xu, \& Li]{lvSystemDesignImproved2020}
Lv, G., Chen, Y., Feng, H., Xu, Z., \& Li, Q., 2020.
\newblock System {{Design}} for an {{Improved SPIDER Imager}}, in {\em
  Proceedings of the 6th {{China High Resolution Earth Observation Conference}}
  ({{CHREOC}} 2019)\/}, Lecture {{Notes}} in {{Electrical Engineering}}, pp.
  241--259, {Springer}, {Singapore}.

\bibitem[Mukherjee et~al.(2020)Mukherjee, Dittmer, Shumaylov, Lunz, {\"O}ktem,
  \& Sch{\"o}nlieb]{mukherjeeLearnedConvexRegularizers2020}
Mukherjee, S., Dittmer, S., Shumaylov, Z., Lunz, S., {\"O}ktem, O., \&
  Sch{\"o}nlieb, C.-B., 2020.
\newblock Learned convex regularizers for inverse problems, {\it ArXiv
  e-print\/}.

\bibitem[Mukherjee et~al.(2021{\natexlab{a}})Mukherjee, Carioni, {\"O}ktem, \&
  Sch{\"o}nlieb]{mukherjeeEndtoendReconstructionMeets2021}
Mukherjee, S., Carioni, M., {\"O}ktem, O., \& Sch{\"o}nlieb, C.-B.,
  2021{\natexlab{a}}.
\newblock End-to-end reconstruction meets data-driven regularization for
  inverse problems, {\it ArXiv e-print\/}.

\bibitem[Mukherjee et~al.(2021{\natexlab{b}})Mukherjee, {\"O}ktem, \&
  Sch{\"o}nlieb]{mukherjeeAdversariallyLearnedIterative2021}
Mukherjee, S., {\"O}ktem, O., \& Sch{\"o}nlieb, C.-B., 2021{\natexlab{b}}.
\newblock Adversarially {{Learned Iterative Reconstruction}} for {{Imaging
  Inverse Problems}}, in {\em Scale {{Space}} and {{Variational Methods}} in
  {{Computer Vision}}\/}, vol. 12679, pp. 540--552, eds Elmoataz, A., Fadili,
  J., Qu{\'e}au, Y., Rabin, J., \& Simon, L., {Springer International
  Publishing}, {Cham}.

\bibitem[Nelson et~al.(2019{\natexlab{a}})Nelson, Pillepich, Springel, Pakmor,
  Weinberger, Genel, Torrey, Vogelsberger, Marinacci, \&
  Hernquist]{nelsonFirstResultsTNG502019}
Nelson, D., Pillepich, A., Springel, V., Pakmor, R., Weinberger, R., Genel, S.,
  Torrey, P., Vogelsberger, M., Marinacci, F., \& Hernquist, L.,
  2019{\natexlab{a}}.
\newblock First {{Results}} from the {{TNG50 Simulation}}: {{Galactic}}
  outflows driven by supernovae and black hole feedback, {\it Monthly Notices
  of the Royal Astronomical Society\/}, {\bf 490}(3), 3234--3261.

\bibitem[Nelson et~al.(2019{\natexlab{b}})Nelson, Springel, Pillepich,
  {Rodriguez-Gomez}, Torrey, Genel, Vogelsberger, Pakmor, Marinacci,
  Weinberger, Kelley, Lovell, Diemer, \&
  Hernquist]{nelsonIllustrisTNGSimulationsPublic2019}
Nelson, D., Springel, V., Pillepich, A., {Rodriguez-Gomez}, V., Torrey, P.,
  Genel, S., Vogelsberger, M., Pakmor, R., Marinacci, F., Weinberger, R.,
  Kelley, L., Lovell, M., Diemer, B., \& Hernquist, L., 2019{\natexlab{b}}.
\newblock The {{IllustrisTNG}} simulations: Public data release, {\it
  Computational Astrophysics and Cosmology\/}, {\bf 6}(1), 2.

\bibitem[Onose et~al.(2016)Onose, Carrillo, Repetti, McEwen, Thiran, Pesquet,
  \& Wiaux]{onoseScalableSplittingAlgorithms2016}
Onose, A., Carrillo, R.~E., Repetti, A., McEwen, J.~D., Thiran, J.-P., Pesquet,
  J.-C., \& Wiaux, Y., 2016.
\newblock Scalable splitting algorithms for big-data interferometric imaging in
  the {{SKA}} era, {\it Monthly Notices of the Royal Astronomical Society\/},
  {\bf 462}(4), 4314--4335.

\bibitem[Pan \& Betcke(2022)]{panLearningInvisiblePhotoacoustic2022}
Pan, B. \& Betcke, M.~M., 2022.
\newblock On {{Learning}} the {{Invisible}} in {{Photoacoustic Tomography}}
  with {{Flat Directionally Sensitive Detector}}, {\it ArXiv e-print\/}.

\bibitem[Pereyra et~al.(2015)Pereyra, {Bioucas-Dias}, \&
  Figueiredo]{pereyraMaximumaposterioriEstimationUnknown2015}
Pereyra, M., {Bioucas-Dias}, J.~M., \& Figueiredo, M. A.~T., 2015.
\newblock Maximum-a-posteriori estimation with unknown regularisation
  parameters, in {\em 2015 23rd {{European Signal Processing Conference}}
  ({{EUSIPCO}})\/}, pp. 230--234, {IEEE}, {Nice}.

\bibitem[Pillepich et~al.(2019)Pillepich, Nelson, Springel, Pakmor, Torrey,
  Weinberger, Vogelsberger, Marinacci, Genel, {van der Wel}, \&
  Hernquist]{pillepichFirstResultsTNG502019}
Pillepich, A., Nelson, D., Springel, V., Pakmor, R., Torrey, P., Weinberger,
  R., Vogelsberger, M., Marinacci, F., Genel, S., {van der Wel}, A., \&
  Hernquist, L., 2019.
\newblock First {{Results}} from the {{TNG50 Simulation}}: {{The}} evolution of
  stellar and gaseous disks across cosmic time, {\it Monthly Notices of the
  Royal Astronomical Society\/}, {\bf 490}(3), 3196--3233.

\bibitem[Pratley \& Mcewen(2021)]{pratleySparseImageReconstruction2021}
Pratley, L. \& Mcewen, J., 2021.
\newblock Sparse {{Image Reconstruction}} for the {{SPIDER Optical
  Interferometric Telescope}}:, in {\em Proceedings of the 9th {{International
  Conference}} on {{Photonics}}, {{Optics}} and {{Laser Technology}}\/}, pp.
  104--109, {SCITEPRESS - Science and Technology Publications}.

\bibitem[Pratley et~al.(2018)Pratley, McEwen, {d'Avezac}, Carrillo, Onose, \&
  Wiaux]{pratleyRobustSparseImage2018}
Pratley, L., McEwen, J.~D., {d'Avezac}, M., Carrillo, R.~E., Onose, A., \&
  Wiaux, Y., 2018.
\newblock Robust sparse image reconstruction of radio interferometric
  observations with purify, {\it Monthly Notices of the Royal Astronomical
  Society\/}, {\bf 473}(1), 1038--1058.

\bibitem[Pratley et~al.(2019)Pratley, McEwen, {d'Avezac}, Cai, {Perez-Suarez},
  Christidi, \& Guichard]{pratleyDistributedParallelSparse2019}
Pratley, L., McEwen, J.~D., {d'Avezac}, M., Cai, X., {Perez-Suarez}, D.,
  Christidi, I., \& Guichard, R., 2019.
\newblock Distributed and parallel sparse convex optimization for radio
  interferometry with {{PURIFY}}, {\it ArXiv e-print\/}.

\bibitem[Putzky \& Welling(2017)]{putzkyRecurrentInferenceMachines2017}
Putzky, P. \& Welling, M., 2017.
\newblock Recurrent {{Inference Machines}} for {{Solving Inverse Problems}},
  {\it ArXiv e-print\/}.

\bibitem[Ronneberger et~al.(2015)Ronneberger, Fischer, \&
  Brox]{ronnebergerUNetConvolutionalNetworks2015}
Ronneberger, O., Fischer, P., \& Brox, T., 2015.
\newblock U-{{Net}}: {{Convolutional Networks}} for {{Biomedical Image
  Segmentation}}, in {\em Medical {{Image Computing}} and {{Computer-Assisted
  Intervention}} \textendash{} {{MICCAI}} 2015\/}, vol. 9351, pp. 234--241, eds
  Navab, N., Hornegger, J., Wells, W.~M., \& Frangi, A.~F., {Springer
  International Publishing}, {Cham}.

\bibitem[Ryu et~al.(2019)Ryu, Liu, Wang, Chen, Wang, \&
  Yin]{ryuPlugandPlayMethodsProvably2019}
Ryu, E.~K., Liu, J., Wang, S., Chen, X., Wang, Z., \& Yin, W., 2019.
\newblock Plug-and-{{Play Methods Provably Converge}} with {{Properly Trained
  Denoisers}}, {\it ArXiv e-print\/}.

\bibitem[Sault \& Conway(1999)]{saultMultiFrequencySynthesis1999}
Sault, {\relax RJ}. \& Conway, {\relax JE}., 1999.
\newblock Multi-frequency synthesis, in {\em Synthesis Imaging in Radio
  Astronomy {{II}}\/}, vol. 180, p. 419.

\bibitem[Starck et~al.(2010)Starck, Murtagh, \&
  Fadili]{starckSparseImageSignal2010}
Starck, J.-L., Murtagh, F., \& Fadili, J.~M., 2010.
\newblock {\it Sparse {{Image}} and {{Signal Processing}}: {{Wavelets}},
  {{Curvelets}}, {{Morphological Diversity}}\/}, {Cambridge University Press},
  1st edn.

\bibitem[Su et~al.(2017)Su, Scott, Ogden, Thurman, Kendrick, Duncan, Yu, \&
  Yoo]{suExperimentalDemonstrationInterferometric2017}
Su, T., Scott, R.~P., Ogden, C., Thurman, S.~T., Kendrick, R.~L., Duncan, A.,
  Yu, R., \& Yoo, S. J.~B., 2017.
\newblock Experimental demonstration of interferometric imaging using photonic
  integrated circuits, {\it Optics Express\/}, {\bf 25}(11), 12653.

\bibitem[Su et~al.(2018)Su, Liu, Badham, Thurman, Kendrick, Duncan, Wuchenich,
  Ogden, Chriqui, Feng, Chun, Lai, \& Yoo]{suInterferometricImagingUsing2018}
Su, T., Liu, G., Badham, K.~E., Thurman, S.~T., Kendrick, R.~L., Duncan, A.,
  Wuchenich, D., Ogden, C., Chriqui, G., Feng, S., Chun, J., Lai, W., \& Yoo,
  S. J.~B., 2018.
\newblock Interferometric imaging using {{Si}} {\textsubscript{3}} {{N}}
  {\textsubscript{4}} photonic integrated circuits for a {{SPIDER}} imager,
  {\it Optics Express\/}, {\bf 26}(10), 12801.

\bibitem[Taylor et~al.(1999)Taylor, Carilli, Perley, \&
  Observatory]{taylorSynthesisImagingRadio1999}
eds Taylor, G.~B., Carilli, C.~L., Perley, R.~A., \& Observatory, N. R.~A.,
  1999.
\newblock {\it Synthesis Imaging in Radio Astronomy. 2: {{Sixth NRAO}}/{{NMIMT
  Synthesis Imaging Summer School}}, 17 - 23 {{June}} 1998 / Ed. by {{G}}.
  {{B}}. {{Taylor}}, {{C}}. {{L}}. {{Carilli}}, and {{R}}. {{A}}.
  {{Perley}}\/}, no. Vol. 180 in Conference Series / {{Astronomical Society}}
  of the {{Pacific}}, {Astronomical Society of the Pacific}, {San Francisco,
  Calif}.

\bibitem[Terris et~al.(2019)Terris, Abdulaziz, Dabbech, Jiang, Repetti,
  Pesquet, \& Wiaux]{terrisDeepPostProcessingSparse2019}
Terris, M., Abdulaziz, A., Dabbech, A., Jiang, M., Repetti, A., Pesquet, J.-C.,
  \& Wiaux, Y., 2019.
\newblock Deep {{Post-Processing}} for {{Sparse Image Deconvolution}}, {\it
  Signal Processing with Adaptive Sparse Structured Representations (SPARS)
  workshop\/}, p.~3.

\bibitem[Terris et~al.(2022)Terris, Dabbech, Tang, \&
  Wiaux]{terrisImageReconstructionAlgorithms2022}
Terris, M., Dabbech, A., Tang, C., \& Wiaux, Y., 2022.
\newblock Image reconstruction algorithms in radio interferometry: From
  handcrafted to learned regularization denoisers, {\it ArXiv e-print\/}.

\bibitem[Trent(2020)]{trentArchitectureInspiredSolvers2020}
Trent, D.~B., 2020.
\newblock {\it Architecture {{Inspired Solvers}}: {{Multi-scale Learnt
  Methods}}\/}, {{MSc}}. {{Thesis}}, University College London, {London}.

\bibitem[Venkatakrishnan et~al.(2013)Venkatakrishnan, Bouman, \&
  Wohlberg]{venkatakrishnanPlugandPlayPriorsModel2013}
Venkatakrishnan, S.~V., Bouman, C.~A., \& Wohlberg, B., 2013.
\newblock Plug-and-{{Play}} priors for model based reconstruction, in {\em 2013
  {{IEEE Global Conference}} on {{Signal}} and {{Information Processing}}\/},
  pp. 945--948, {IEEE}, {Austin, TX, USA}.

\bibitem[Wang et~al.(2004)Wang, Bovik, Sheikh, \&
  Simoncelli]{wangImageQualityAssessment2004}
Wang, Z., Bovik, A., Sheikh, H., \& Simoncelli, E., 2004.
\newblock Image {{Quality Assessment}}: {{From Error Visibility}} to
  {{Structural Similarity}}, {\it IEEE Transactions on Image Processing\/},
  {\bf 13}(4), 600--612.

\bibitem[Xu et~al.(2012)Xu, Yu, Mou, Zhang, Hsieh, \&
  Wang]{xuLowdoseXrayCT2012}
Xu, Q., Yu, H., Mou, X., Zhang, L., Hsieh, J., \& Wang, G., 2012.
\newblock Low-dose {{X-ray CT}} reconstruction via dictionary learning, {\it
  IEEE Transactions on Medical Imaging\/}, {\bf 31}(9), 1682--1697.

\bibitem[Yang et~al.(2017)Yang, Sun, Li, \& Xu]{yangADMMNetDeepLearning2017}
Yang, Y., Sun, J., Li, H., \& Xu, Z., 2017.
\newblock {{ADMM-Net}}: {{A Deep Learning Approach}} for {{Compressive Sensing
  MRI}}, {\it ArXiv e-print\/}.

\bibitem[Yi \& Babyn(2018)]{yiSharpnessAwareLowDoseCT2018}
Yi, X. \& Babyn, P., 2018.
\newblock Sharpness-{{Aware Low-Dose CT Denoising Using Conditional Generative
  Adversarial Network}}, {\it Journal of Digital Imaging\/}, {\bf 31}(5),
  655--669.

\bibitem[Zernike(1938)]{zernikeConceptDegreeCoherence1938}
Zernike, F., 1938.
\newblock The concept of degree of coherence and its application to optical
  problems, {\it Physica\/}, {\bf 5}(8), 785--795.

\bibitem[Zheng et~al.(2020)Zheng, Gao, Zhang, \&
  Xing]{zhengDualdomainDeepLearningbased2020}
Zheng, A., Gao, H., Zhang, L., \& Xing, Y., 2020.
\newblock A dual-domain deep learning-based reconstruction method for fully
  {{3D}} sparse data helical {{CT}}, {\it Physics in Medicine \& Biology\/},
  {\bf 65}(24), 245030.

\end{thebibliography}








\bsp	
\label{lastpage}
\end{document}